\documentclass{article}
\makeatletter
\def\input@path{{}}
\makeatother
\usepackage{aibl_arxiv}
\usepackage[T1]{fontenc}
\usepackage[utf8]{inputenc}
\usepackage{amsmath,amssymb,mathtools,amsthm}
\usepackage{graphicx,booktabs,array,tabularx,multirow}
\usepackage{tikz}
\usetikzlibrary{arrows.meta,positioning,calc}
\usepackage{microtype,needspace,placeins,float}
\usepackage{xurl,bookmark}
\hypersetup{pdftitle={Complete Donor-Set Guidance for Frozen Diffusion Docking Models},pdfauthor={Hyunwoo Kim, Insoo Park, Ziseok Lee, Wooyeol Lee, Yoorim Gang, Kyungsu Kim}}
\newcolumntype{Y}{>{\raggedright\arraybackslash}X}
\newcolumntype{P}[1]{>{\raggedright\arraybackslash}p{#1}}
\newcommand{\ang}{\,\text{\AA}}

\newcommand{\Sel}{\operatorname{Sel}}

\newcommand{\CorrespondingEmail}{kyskim@snu.ac.kr}
\begin{document}
\aiblmaketitle
  {Complete Donor-Set Guidance for Frozen Diffusion Docking Models}
  {Hyunwoo Kim\textsuperscript{1,*}, Insoo Park\textsuperscript{1,*}, Ziseok Lee\textsuperscript{1},\\[2pt]
   Wooyeol Lee\textsuperscript{1}, Yoorim Gang\textsuperscript{1}, Kyungsu Kim\textsuperscript{1,\textdagger}}
  {\textsuperscript{1}Seoul National University\\[2pt]
   {\small \textsuperscript{*}Equal contribution.\quad \textsuperscript{\textdagger}Corresponding author:
   \texttt{\CorrespondingEmail}}}
  {Metalloprotein docking requires a ligand pose to satisfy both global pocket fit and the identity and geometry of metal-coordinating atoms. General-purpose diffusion docking models do not explicitly condition generation on externally specified complete ligand donor sets, and their predicted poses can exhibit incorrect metal--donor relationships. Metal-aware models address coordination during training and generation; here we ask whether external donor information can correct already-trained general-purpose models without retraining. We propose Complete-Identity Contract Guidance (CICG), which applies a supplied complete donor set through translation, rotation, and torsion updates, and Historical Contract Transfer (HCT), which supplies donor identities from earlier complexes at compatible receptor sites. We compare HCT with trajectory-local controls retaining the historical donor count with or without elemental composition. On 86 assessable HCT-covered queries, HCT-guided CICG increases frozen-confidence-selected joint success from 21.5\% to 36.6\% on SurfDock and from 25.0\% to 34.9\% on DiffDock-L. The joint criterion requires correct global ligand placement, complete native donor recovery, coordination geometry, preserved stereochemistry, and absence of fatal receptor clashes. Composition-matched guidance reaches 33.14\% and 34.30\%, leaving the additional benefit of fixed historical membership unresolved. Our results therefore support donor-set guidance as a conditional correction for frozen docking models, while identifying donor-source reliability and physical-quality preservation as the main barriers to broader use.}
  {\aiblresource{Github}{https://github.com/pe7ace/CICG}}

\section{Introduction}
A ligand binding a metalloprotein must satisfy two coupled requirements: its overall conformation must fit the protein pocket, and appropriate ligand atoms must coordinate the metal. Position, orientation, and torsion angles are continuous variables, whereas the identities of coordinating atoms form a discrete set. Two possible sets can share both donor count and elemental composition yet constrain different parts of the same molecule. For example, $\{O_a,O_b\}$ and $\{O_a,O_c\}$ each contain two oxygen donors but specify different atom memberships.

General-purpose diffusion dockers \citep{corso2023diffdock,cao2025surfdock} do not explicitly condition generation on an externally specified complete ligand donor set, so pocket fit can coexist with an incorrect metal--donor relationship. Metal-aware methods, including MetalloDock, already model coordination during training and generation \citep{zhang2026metallodock}. We ask whether supplied donor information can correct frozen general-purpose dockers without retraining. 

A practical motivation comes from analogue optimization of metalloprotein inhibitors. \citet{bonardi2020tails} evaluated 32 benzenesulfonamide derivatives with different tail combinations; five hCA II co-crystal structures showed a conserved zinc-binding arrangement alongside distinct tail interactions. Similarly, \citet{lee2016cryptic} combined crystallography and solution NMR to guide optimization of hydroxamate-containing LpxC inhibitors by characterizing alternative bound conformations. These studies motivate a setting in which related compounds provide a plausible metal-binding group while the full pose of a new analogue remains to be predicted.

We propose \textbf{Complete-Identity Contract Guidance (CICG)}, which reduces metal--donor distance residuals through translation, rotation, and torsion updates at prespecified late-stage sampling hooks. Donor assignments and pretrained generator/confidence parameters stay fixed. We use the term \emph{donor contract} to refer to a complete ligand-side donor set. \textbf{Historical Contract Transfer (HCT)} supplies mapped identities from earlier complexes at compatible receptor sites, without query-native donor labels or reference-coordinate targets. When no reference passes its transfer rules, HCT abstains and retains unmodified docking. It covers 62/304 original-roster queries (20.39\%).

Historical transfer and trajectory-local nearest-donor selection construct alternative inputs to the same CICG operator. The nearest policies use the donor count inferred from the historical reference, optionally with its N/O/S composition, but choose atom identities separately for each trajectory. Native-donor guidance is a separate diagnostic; post-hoc donor-aware selection tests use of historical information without coordinate modification. We evaluate a joint criterion requiring correct global ligand placement, complete native donor recovery, coordination geometry, preserved stereochemistry, and absence of fatal receptor clashes.

\begin{figure}[!htbp]
\centering
\includegraphics[width=\linewidth]{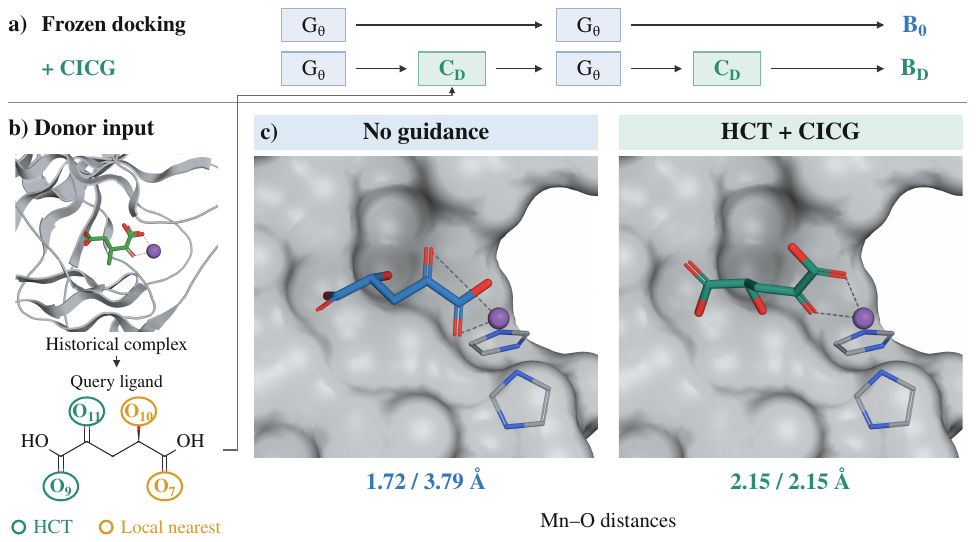}
\caption{\textbf{Supplied-donor correction of a frozen docking model.}
(a) CICG applies donor-conditioned corrections $C_D$ after late sampling updates of the frozen generator $G_\theta$.
(b) The query-ligand graph shows alternative donor sets supplied by HCT (green) and trajectory-local nearest selection (orange). Both enter the same correction operator; the ribbon view shows a historical reference complex.
(c) Frozen-confidence-selected SurfDock poses illustrate the metal--donor geometry. Gray surfaces show the receptor pocket; blue and green ligand carbons distinguish no guidance and HCT + CICG. Dashed connectors show Mn--donor distances; values are ordered Mn--O9 / Mn--O11. Structural details are given in Appendix~\ref{app:figure1_provenance}.}

\label{fig:overview}
\end{figure}

Our contribution is a common correction interface and a controlled comparison of donor-information policies and coordinate versus selection-only use. On the evaluated HCT-covered queries, donor guidance improves joint success on two frozen backbones.
Figure~\ref{fig:overview} shows the donor-set inputs, how they integrate into frozen sampling, and a structural comparison of the resulting poses.

\section{Related work}
\paragraph{Docking with metal and structural information.} DiffDock models ligand translation, rotation, and torsions with diffusion, while SurfDock incorporates protein-surface representations into this geometric framework \citep{corso2023diffdock,cao2025surfdock}. Metal-specific docking methods incorporate coordination information through different mechanisms. AutoDock4Zn modifies the docking potential for zinc coordination \citep{santosmartins2014autodockzn}, whereas GM-DockZn generates poses by matching potential coordinating atoms to zinc coordination geometries and evaluates both coordinating-atom identity and coordination number alongside pose RMSD \citep{wang2020gmdockzn}. Metalloprotein Bias Docking extends atom-specific interaction biases to multiple metal classes \citep{clemente2024mbd}, while MetalloDock learns metal-aware donor and pose generation directly \citep{zhang2026metallodock}. HCT instead transfers ligand donor membership from compatible historical complexes, and CICG uses that external information to modify a frozen generator without retraining it.

\paragraph{Guidance of pretrained diffusion models.} GeoDirDock guides translational, rotational, and torsional variables and supports template-based docking through geometric information transferred on a maximum common substructure \citep{minan2025geodirdock}. HCT differs in transferring donor membership rather than reference ligand coordinates or torsion targets. More generally, Universal Guidance, Projected Diffusion, and Stein Diffusion Guidance modify pretrained generative sampling through external objectives, constrained updates, or posterior correction \citep{bansal2024universal, christopher2024pdm, nguyen2026stein}. Force-field guidance likewise introduces physical information during protein-conditioned molecular generation without retraining the generator \citep{lai2026forcefield}. CICG uses a local kinematic correction to compare alternative donor-information policies within frozen docking models.

\paragraph{Generation, selection, and validity.} Coordinate correction changes both candidate geometry and the inputs seen by the frozen confidence model, so candidate availability and selected-pose success need not change together. PoseBusters shows that low RMSD alone does not ensure physical validity \citep{buttenschoen2024posebusters}, and interaction-recovery studies similarly distinguish pose placement from recovery of protein contacts \citep{errington2025interaction}. Our joint endpoint additionally evaluates complete donor membership and coordination geometry under a whole-ligand mapping. More broadly, predictive accuracy and downstream decision quality need not coincide \citep{elmachtoub2022spo, mandi2022decision}. We therefore analyze donor-set correctness, geometric realization, candidate availability, final selection, and candidate-level physical quality separately.

Table~\ref{tab:related_scope} summarizes the information sources
and mechanisms of these approaches.

\begin{table}[t]
\centering
\caption{Information sources and mechanisms of related approaches and CICG/HCT.}
\label{tab:related_scope}
\begingroup
\setlength{\tabcolsep}{3pt}
\renewcommand{\arraystretch}{1.10}
\normalsize
\fitbox{\begin{tabularx}{\linewidth}{@{}P{80pt}P{108pt}Y@{}}
\toprule
Method & Information used & Mechanism \\
\midrule
GeoDirDock \citep{minan2025geodirdock}
& Geometric priors; template coordinates and torsions
& Guidance in translation, rotation, and torsion spaces. \\
\addlinespace
MetalloDock \citep{zhang2026metallodock}
& Metal pocket and ligand; optional coordination-atom input
& Learned metal-aware donor prediction and pose generation. \\
\addlinespace
MBD \citep{clemente2024mbd}
& Atom-specific metal interaction biases
& Bias terms incorporated into the docking potential. \\
\addlinespace
Universal Guidance / PDM
\citep{bansal2024universal,christopher2024pdm}
& External objectives / constraint sets
& Guided sampling / projection onto constraints. \\
\addlinespace
Force-field guidance \citep{lai2026forcefield}
& Force-field energies
& Energy guidance during protein-conditioned molecular generation. \\
\addlinespace
CICG / HCT (this work)
& Supplied complete donor sets; historical complexes for HCT
& Donor-distance correction of frozen docking models through
translation, rotation, and torsion updates. \\
\bottomrule
\end{tabularx}}
\endgroup
\end{table}

\section{Complete-Identity Contract Guidance}
\label{sec:method}
\subsection{A donor-set condition for a frozen generator}
A query is $X=(P,L,s_m,\mathbf z_m)$, with receptor $P$, chemically specified ligand graph $L$, metal element $s_m$, and metal coordinate $\mathbf z_m$. A complete donor set (or donor contract) $D\subseteq V(L)$ specifies all ligand atoms intended to coordinate the selected metal; the final pose may realize a different membership. We write $\widehat D$ for the historical assignment and $D^\star$ for the native set used in evaluation and the diagnostic oracle. Atom indices are mapped to the query ligand before sampling.

A pose $y$ specifies the ligand heavy-atom coordinates in the receptor frame. A bank $\mathcal B=(y_k)_k$ is an indexed candidate collection, with 40 poses in the main experiments and associated confidence scores; $k$ is the recorded sample identifier, not confidence rank.

Let $\mathcal B_0=G_\theta(X;\xi)$ be the unmodified bank and $\mathcal B_D=\mathcal A(G_\theta,X,D;\xi)$ the guided bank when all trajectories share $D$. Here $\xi$ denotes the initial randomness and subsequent trajectory noise. Historical and native-donor policies share one set across trajectories. For nearest-donor policies, the set is trajectory-specific; we denote the resulting banks by
$\mathcal B_{\rm near}$ and $\mathcal B_{\rm comp}$. Each trajectory holds its own assigned set fixed throughout correction. Within each backbone, all conditions use the same pretrained generator
parameters $\theta$ and confidence parameters $\eta$.
All guided conditions use the same CICG operator.

\subsection{Coordinate update}
\label{sec:operator}
At an intervention, let $q=(\mathbf t,\boldsymbol\omega,\varphi_1,\ldots,\varphi_{n_{\rm rot}})$ contain $6+n_{\rm rot}$ local coordinates. Here
$\mathbf t\in\mathbb R^3$ is translation in \AA, $\boldsymbol\omega\in\mathbb R^3$ is a local rotation vector, and the $n_{\rm rot}$ torsion coordinates correspond to the adjusted rotatable bonds. Angular coordinates are in radians. The backbone's coordinate-update routine gives atom positions $\mathbf x_i(q)$.
Define $d_i^{\rm tar}:=d^{\rm tar}_{s_m,e_i}$ from the fixed metal--element table (Table~\ref{tab:targets}), where $e_i$ is the element of atom $i$. The normalized radial residual is
\begin{equation}
 r_i(q)=
 \frac{\|\mathbf x_i(q)-\mathbf z_m\|_2-d_i^{\rm tar}}{\sigma},
 \qquad i\in D,\qquad \sigma=0.10\ang.
 \label{eq:residual}
\end{equation}
We stack these residuals in a fixed donor-index order as
$r(q)=(r_i(q))_{i\in D}\in\mathbb R^{|D|}$ and locally reduce $\tfrac12\|r(q)\|_2^2$. The quantity $\sigma$ is a fixed normalization scale.

At iteration $j$, $J_j\in\mathbb R^{|D|\times(6+n_{\rm rot})}$ approximates $\left.\partial r/\partial q\right|_{q_j}$ by central finite differences with step $10^{-5}$ in the stored coordinates.
The coordinate metric is $M=\operatorname{diag}(I_6,8I_{n_{\rm rot}})$, where $I_m$ is the $m$-dimensional identity matrix. It assigns unit weight to translation/rotation changes and weight eight to torsion
changes in this numerical parameterization. The trial direction is
\begin{equation}
 \Delta q_j=-M^{-1/2}\big(J_jM^{-1/2}\big)^+r(q_j).
 \label{eq:update}
\end{equation}
The superscript $+$ denotes the SVD-based pseudoinverse with relative singular-value cutoff $10^{-10}$. This gives the minimum-metric-norm least-squares correction in the retained singular subspace. Accepted updates are followed by relinearization.

For each $\alpha\in\{1,2^{-1},\ldots,2^{-15}\}$, the trial state is $q_{\rm trial}=\operatorname{Retr}_{q_j}(\alpha\Delta q_j)$, using the backbone's kinematic retraction specified in Appendix~\ref{app:implementation}. A trial is accepted only if its coordinates are finite, its residual norm decreases, and its heavy-atom Cartesian RMS displacement from the current iterate does not exceed $0.25\ang$. There are at most five relinearizations per intervention, with early stopping at $\|r\|_2\le10^{-3}$. An unsuccessful trial leaves the current iterate unchanged. The reference SurfDock implementation applies the operator after steps 18 and 19 of its 20-step sampler (zero-based indexing), to all 40 trajectories. A timing ablation of this operator places both calls after the final native update (Section~\ref{sec:timing}). The displacement cap applies to each accepted update; receptor clashes are evaluated after sampling.

\subsection{Historical Contract Transfer}
\label{sec:hct}
HCT uses a collection $\mathcal H$ of earlier complexes to return one query donor set or abstain:
\begin{equation}
 \widehat D=S_{\rm HCT}(X,\mathcal H)\in\mathcal C(L)\cup\{\bot\}.
 \label{eq:supplier}
\end{equation}
Here $\mathcal C(L)$ is the set of admissible ligand donor sets and $\bot$ denotes abstention; $S_{\rm HCT}$ returns a contract, not a pose. The reference collection contains 2,344 usable records retained from 2,423 raw records, released before 2023 and before the corresponding query. The strict matching rule requires the same metal; at least 90\% sequence identity and 90\% coverage in each alignment direction; matching aligned protein first shells defined within $3\ang$ of the metal; an earlier reference release; different ligand components and canonical graphs; and agreement of mapped donor sets.

Two extensions increase coverage. Donor-core matching replaces a global maximum common substructure with the radius-two graph neighborhood of the reference donor atoms. Graph-deduplicated consensus counts repeated observations with the same canonical ligand graph as a single vote. These rules and their outputs were fixed before comparison with query-native donor labels. The historical coordinates are used to identify reference donors and compare receptor sites, but only the mapped donor identities enter the correction. Appendix~\ref{app:historical} details the resulting coverage and precision.

On queries receiving a contract, the guided bank replaces the unmodified bank. The frozen-confidence selector $\Sel_\eta(\mathcal B)\in\mathcal B$ returns the highest-scoring pose:
\begin{equation}
 \Pi(X)=
 \begin{cases}
 \Sel_\eta(\mathcal B_{\widehat D}),&\widehat D\ne\bot,\\
 \Sel_\eta(\mathcal B_0),&\widehat D=\bot,
 \end{cases}
 \qquad
 \Sel_\eta(\mathcal B)=\arg\max_{y\in\mathcal B}s_\eta(y).
 \label{eq:policy}
\end{equation}
The score $s_\eta(y)$ is computed for each candidate using the original frozen confidence model, with ties resolved by its original ranking implementation. All supplied contracts are applied, while queries without a contract retain the unmodified bank.

\subsection{Trajectory-specific nearest-donor policies}
\label{sec:nearest}
Nearest-K uses the donor count supplied by the historical reference, $K_H=|\widehat D|$, but selects a donor set independently for each trajectory. Let $\mathbf x_i^{\rm pre}$ denote the coordinates of atom $i$ immediately before the first scheduled correction, and let $V_{\rm eligible}(L,s_m)$ be the supported nitrogen, oxygen, and sulfur atoms for the target metal. The count-only policy selects
\begin{equation}
D_{\rm near}
=
\arg\min_{\substack{
D\subseteq V_{\rm eligible}\\
|D|=K_H
}}
\sum_{i\in D}
\|\mathbf x_i^{\rm pre}-\mathbf z_m\|_2    
\end{equation}

Distance ties are resolved by input atom index. Eligibility is oxygen for calcium; nitrogen or oxygen for magnesium; and nitrogen, oxygen, or sulfur for cobalt, iron, manganese, and zinc. The chosen set remains fixed throughout all correction calls and their finite-difference, line-search, and relinearization calculations. Historical information still supplies both $K_H$ and the applicable query set.

The \emph{composition-matched Nearest-$K$} control additionally preserves the historical count $n_e(\widehat D)$ of each element $e\in\{\mathrm N,\mathrm O,\mathrm S\}$. Writing $V_e$ for eligible atoms of element $e$, it selects
\begin{equation}
\begin{aligned}
D_{\rm comp}
&=
\arg\min_{D\subseteq V_{\rm eligible}}
\sum_{i\in D}
\|\mathbf x_i^{\rm pre}-\mathbf z_m\|_2,\\
&\text{subject to}\quad
|D\cap V_e|=n_e(\widehat D)
\quad
\text{for each }e\in\{\mathrm N,\mathrm O,\mathrm S\}.
\end{aligned}
\end{equation}
This holds the target-distance composition fixed while changing which atoms receive the constraints. For SurfDock, $\mathbf x^{\rm pre}$ is taken after native
update 18, and the selected set is retained through correction after update 19. For DiffDock-L, it is taken after native update 17 and retained through correction after update 18. Both controls inherit historical scope and cardinality; composition matching additionally inherits elemental quotas, which may themselves be incorrect.

\subsection{Donor-aware selection and coordinate-bank ablation}
\label{sec:scoring}
We apply one deterministic donor-aware rule to both the unmodified and historical-guided 40-pose banks to compare coordinate construction under a common selection policy. For a pose $y$, let $N_{K_H}(y)$ be its nearest $K_H$ ligand heavy atoms and $\mathcal G_L$ the chirality-preserving automorphisms of the input ligand graph. The overlap deficit is
\begin{equation}
e(y)
=
K_H-
\max_{\pi\in\mathcal G_L}
\left|
N_{K_H}(y)\cap\pi(\widehat D)
\right|
 \label{eq:scoring}
\end{equation}
The deficit $e(y)\in\{0,\ldots,K_H\}$ is zero when the nearest $K_H$ ligand heavy atoms agree with the complete historical donor set under an allowed input-graph permutation. This nearest-atom search includes all ligand heavy-atom elements, whereas the contract-construction policies use metal-specific eligible atoms.
Candidates are ordered by increasing $e(y)$, decreasing frozen-model confidence for that pose, and increasing sample identifier. Selection uses the input ligand graph, historical donor set, and evaluated bank; native labels enter evaluation only. The unchanged rule is applied separately to both banks. We denote the first pose under this ordering by $S(\mathcal B;\widehat D)\in\mathcal B$, with the fixed query inputs and confidence-model parameters suppressed from the notation. The reranking and bank-by-selector comparisons were specified after inspection of the original experiments and are post-hoc analyses. A separately fixed radial/clash rule provides an additional selector diagnostic (Appendix~\ref{app:followup_selector}).

\section{Experimental design}
\label{sec:evaluation}
\paragraph{Data and analysis population.}
The initial roster contains 304 queries: 179 non-overlapping entries from a 400-complex collection and 125 from a separate post-2023 collection. HCT supplies 62 queries in 17 structural/site groups, giving 20.39\% coverage. An additional 38 queries in 22 groups were selected from screened Protein Data Bank collections \citep{berman2000pdb} with fixed metal quotas, without query-native donor labels or docking outcomes. Their structure entries and frozen group identifiers do not overlap the initial cohort. Appendix~\ref{app:historical} gives the selection rules, release ranges, and metal composition.

The stereochemistry checker's input domain covers specified carbon tetrahedral centers and E/Z bonds. An input-only mask excludes 14 ligands with specified non-carbon tetrahedral centers or ligand-internal metals, leaving 86 queries in 33 groups (55 initial and 31 additional). All assessable queries are retained across conditions and seeds. We adopted this pooled, domain-restricted analysis during reporting, after the original outcomes were known. It estimates effects conditional on HCT coverage and input assessability, rather than performance on an independently selected validation cohort. Appendix~\ref{app:fullcohort} reports the original cohort-specific comparisons, including the 62- and 100-query analyses that counted unsupported inputs as failures.

\paragraph{Information and matched execution.}
Each query is sampled at seeds 42 and 43 with 40 trajectories per condition. Historical transfer was fixed before its native-label audit. Native labels subsequently define evaluation and retrospective correctness strata; they do not select which historical contracts are applied. The nearest-donor controls inherit historical cardinality, and the composition-matched control also inherits its elemental quotas.

Within the reference SurfDock policy comparisons, inputs, generator and confidence weights, initial conditions, and corresponding sampling noise are matched. Each guided arm uses the same radial targets, update operator, correction steps, and maximum iteration count. The native-donor oracle supplies atom identities, not native ligand coordinates, distances, or angles as correction targets. Sampling counts and maximum correction iterations are matched; actual runtime can vary. Replication, nearest-donor comparisons, and deterministic reranking were sequential follow-ups to the initial historical-guidance experiment; reranking and label-conditioned diagnostics are post-hoc.

\paragraph{Timing and backbone follow-ups.}
The terminal-only SurfDock control applies two separate calls after native update 19, each with at most five relinearizations. It retains the interleaved condition's donor set; nearest donors are frozen from the post-update-18 pose even when correction is deferred. DiffDock-L uses the v1.1 configuration \citep{corso2024diffdockl} with 19 executed updates, indexed 0--18. Four arms branch from the shared post-update-17 parent, and guided arms are corrected after updates 17 and 18. All six arm pairs have matching saved pre-intervention coordinates; the final update adds no noise under the recorded configuration. All 172 planned blocks are retained, including invalid output slots (Appendix~\ref{app:followup_diffdock}). Both backbones evaluate the same biological cohort.

\paragraph{Joint success.}
We call the evaluation's \emph{clash-free Robust Full} endpoint \emph{joint success}. It requires symmetry-aware heavy-atom root-mean-square deviation (RMSD) below $2.0\ang$ in the receptor frame and a coordination witness: one chirality-preserving whole-ligand mapping under which the $K_\star=|D^\star|$ nearest ligand heavy atoms are the native donors, each radial error is at most $0.50\ang$, and each metal-to-donor vector differs from its native direction by at most $30^\circ$. All heavy-atom elements enter the evaluation's nearest-atom search. Ligand RMSD is minimized separately from the coordination witness.

The same pose must also have finite coordinates, successful molecular sanitization, preserved supported input stereochemistry, and no ligand--receptor heavy-atom separation below $1.2\ang$. These are evaluation criteria, not generation targets. Appendix~\ref{app:evaluator} defines the geometry-valid flag and distinguishes these checks from the full PoseBusters battery. Assigned-set exactness is a separate diagnostic: it asks whether the set fixed at the intervention agrees with the native donor annotation, rather than whether the final coordinates realize that set. We write $R(y)\in\{0,1\}$ for the joint-success indicator of pose $y$, with query-specific evaluation inputs left implicit.

\paragraph{Additional physical checks.}
We aggregate stored PoseBusters checks \citep{buttenschoen2024posebusters} for 27,520 SurfDock poses: no guidance, historical, count-only, and composition-matched guidance. The checklist combines eight internal-quality flags with protein-volume overlap, treating missing flags as failures. The receptor inputs contain protein but no metals, waters, or other cofactors. Checklist compliance and its intersection with joint success are secondary endpoints (Appendix~\ref{app:followup_pb}).

\paragraph{Selection and statistics.}
Top-$k$ success means that at least one of the first $k$ poses in a method's output ranking passes the joint criterion. Sampling conditions use frozen confidence, whereas donor-aware reranking uses Equation~\eqref{eq:scoring}. Top-40 equals availability for a 40-pose bank. Let $\widehat y^a_{n,s}$ be the Top-1 pose selected for query $n$ and seed $s$ by configuration $a$, which specifies both candidate-bank construction and the output-ranking rule. Its joint-success indicator is
$R^a_{n,s}=R(\widehat y^a_{n,s})$. The paired effects are
\begin{equation}
\delta_n^{a,b}
=
\frac12\sum_{s\in\{42,43\}}
\left(R^a_{n,s}-R^b_{n,s}\right),
\qquad
\widehat\Delta^{a,b}
=
\frac1N\sum_{n=1}^{N}\delta_n^{a,b}
 \label{eq:paired}
\end{equation}
Here $N=86$ in the pooled analysis. Positive values favor configuration $a$, and percentage-point differences are reported as $100\widehat\Delta^{a,b}$.
Baseline contrasts assess the benefit of guidance, while historical-versus-nearest contrasts quantify the incremental value of fixed membership. All pooled 86-query contrasts use the same 20,000 draws of the 33 original groups (PCG64, seed 20263001). Each draw retains all paired observations in its sampled groups and weights queries equally, with group multiplicities reflected in the denominator. We report nominal two-sided percentile 95\% intervals. Secondary and post-hoc contrasts are not multiplicity-adjusted. The two seeds and two backbones are repeated evaluations of the same queries, not additional independent biological samples. Appendix~\ref{app:statistics} specifies the common analysis and retains earlier cohort-specific procedures.

\section{Results}
\subsection{Donor guidance improves joint success on two backbones}
Historical guidance increases SurfDock joint Top-1 success from 37/172 to 63/172: +15.12 percentage points [8.00, 24.63]. On the same queries, DiffDock-L increases success from 43/172 to 60/172: +9.88 points [3.09, 17.16] (Table~\ref{tab:policy_backbone_followup}). Composition-matched nearest-donor guidance improves both models. The separate 206-complex native-donor experiment remains in Appendix~\ref{app:oracle}.
\begin{table}[!htbp]
\centering
\caption{Confidence-selected joint success and within-backbone paired effects on the same 86 queries and two seeds. Counts summarize 172 observations per condition. Intervals use the common 33-group bootstrap. The DiffDock-L four-arm follow-up is separate from the 206-complex oracle study.}
\label{tab:policy_backbone_followup}
\fitbox{\begin{tabularx}{\linewidth}{@{}Yrr@{}}\toprule
Policy or paired contrast & SurfDock \citep{cao2025surfdock} \venue{Nat. Methods '25} & DiffDock-L \citep{corso2024diffdockl} \venue{ICLR '24} \\\midrule
No guidance & 37/172 (21.51\%) & 43/172 (25.00\%) \\
Historical donors & 63/172 (36.63\%) & 60/172 (34.88\%) \\
Composition-matched Nearest-$K$ & 57/172 (33.14\%) & 59/172 (34.30\%) \\
Native-donor oracle & 70/172 (40.70\%) & 60/172 (34.88\%) \\\midrule
Historical $-$ no guidance (pp) & $+15.12\ [8.00,24.63]$ & $+9.88\ [3.09,17.16]$ \\
Composition-matched $-$ no guidance (pp) & $+11.63\ [6.12,18.84]$ & $+9.30\ [2.35,17.16]$ \\
Historical $-$ composition-matched (pp) & $+3.49\ [-0.42,8.93]$ & $+0.58\ [-2.03,3.52]$ \\
\bottomrule
\end{tabularx}}
\end{table}

\subsection{Nearest-donor policies recover much of the aggregate benefit}
\label{sec:composition_result}
The historical-minus-composition-matched difference remains uncertain on SurfDock (+3.49 points [$-0.42$, 8.93]) and DiffDock-L (+0.58 [$-2.03$, 3.52]). Count-only guidance reaches 56/172 successes on SurfDock. Native-donor guidance reaches 70/172 and 60/172 on the two backbones. Trajectory-local membership is therefore a competitive use of partial historical information. The DiffDock-L historical and native-donor conditions have the same joint outcomes, although their ligand and identity outcomes differ (Appendix~\ref{app:followup_diffdock}). Full SurfDock contrasts appear in Appendix~\ref{app:followup_policy}.

\subsection{Assignment exactness, realization, and selection are distinct}
\label{sec:assignment_result}
A retrospective native-label audit finds exact assigned sets in 2,688/6,880 composition-matched SurfDock trajectories and 107/172 selected trajectories. Among the latter, 57 pass the joint criterion and 50 fail; all 65 incorrect selected assignments fail (Appendix~\ref{app:final_recovery}). Contract accuracy is therefore insufficient for joint success. The association is evaluated under an endpoint that itself requires native coordination recovery. Historical sets are exact on 69 queries and incorrect on 17; their respective baseline effects are +21.74 points [12.70, 33.13] and $-11.76$ [$-20.00$, 0.00].

\subsection{Coordinate guidance adds benefit under a fixed selector}
\label{sec:fixed_selector}
With the same donor-aware selector applied independently to both banks, replacing the unmodified bank with the historical-guided bank increases joint success from 51/172 to 64/172 (Table~\ref{tab:bank_selector}). The paired gain is +7.56 points [1.11, 14.04], with 21 improvements and eight regressions. Changing the selector on the guided bank alone gives only +0.58 points [$-1.75$, 3.37].
\begin{table}[!htbp]
\centering
\caption{Post-hoc bank-by-selector comparison on 86 queries. Entries count joint Top-1 successes in 172 query--seed observations. Both donor-aware conditions use the same transferred set and ranking rule, evaluated independently on their respective 40-pose banks.}
\label{tab:bank_selector}
\fitbox{\begin{tabular}{lcc}\toprule
Coordinate bank & Frozen confidence & Donor-aware selection \\\midrule
No guidance & 37/172 & 51/172 \\
Historical guidance & 63/172 & 64/172 \\
\bottomrule\end{tabular}}
\end{table}

Under donor-aware selection, ligand success falls from 107/172 to 99/172 ($-4.65$ points [$-8.06$, $-0.68$]) and selected clashes rise from 12/172 to 21/172 (+5.23 [$-0.71$, 11.21]). A separate radial/clash rule yields 48/172 on the baseline bank and 38/172 on the guided bank ($-5.81$ [$-15.13$, 3.80]). The benefit depends on the selection rule and includes downstream changes in both geometry and frozen-model confidence.

\begin{figure}[H]
\centering
\includegraphics[width=\linewidth]{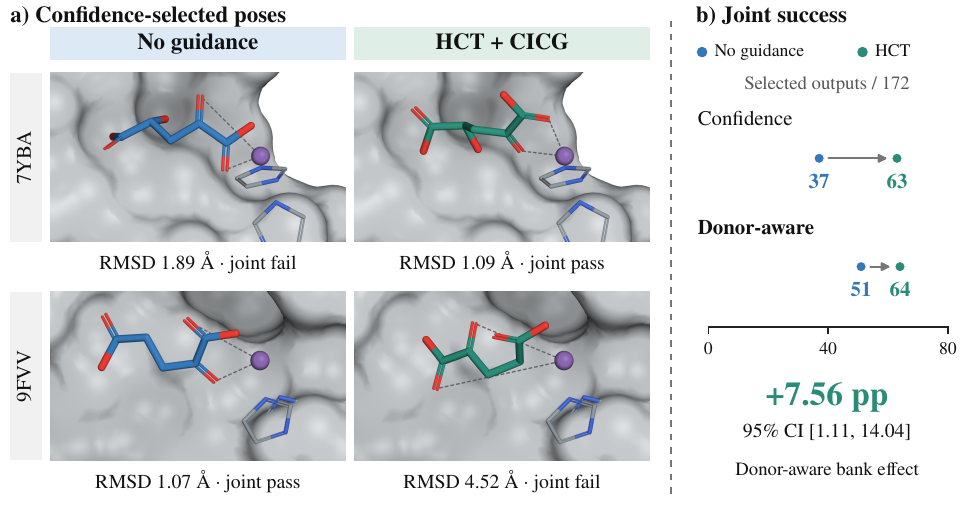}
\caption{\textbf{Structural outcomes and coordinate-bank utility.}
(a) Frozen-confidence-selected SurfDock poses for a recovery (7YBA) and a regression (9FVV). Ligand carbon atoms are blue for no guidance and green for HCT + CICG; oxygen, nitrogen, and the metal are red, blue, and purple. The protein is shown as a gray cutaway surface, and dashed lines connect the transferred donor atoms to the metal. Each pair uses the same camera, and labels report ligand RMSD and joint success.
(b) Selected joint-success counts from the no-guidance and HCT-guided banks under frozen-confidence and post-hoc donor-aware selection on the same 86 queries and two seeds. With donor-aware selection held fixed, success increases from 51 to 64 of 172 observations (+7.56 percentage points; paired 95\% group-bootstrap CI [1.11, 14.04]), comprising 21 recoveries and eight regressions. Banks containing a successful candidate increase from 91 to 96, while selection misses decrease from 40 to 32.}
\label{fig:bank_selection}
\end{figure}

\paragraph{Availability and selection.}
For the fixed donor-aware selector, write $S(\mathcal B)$
for $S(\mathcal B;\widehat D)$, suppressing the historical
contract and fixed query inputs within each pair.
Define the bank-level indicators
\[
\begin{aligned}
A(\mathcal B)
&=\mathbf 1\{\exists y\in\mathcal B:R(y)=1\},\\
Y_S(\mathcal B)
&=R\bigl(S(\mathcal B)\bigr),\\
F_S(\mathcal B)
&=A(\mathcal B)-Y_S(\mathcal B).
\end{aligned}
\]
Thus $F_S=1$ only when a successful candidate exists but the
selected pose fails. Applying the guided-minus-baseline paired
averaging of Equation~\eqref{eq:paired} gives
\[
\Delta Y_S=\Delta A-\Delta F_S.
\]
With donor-aware selection, available successes increase from 91 to 96 banks and selected successes from 51 to 64, while misses decrease from 40 to 32: $13/172=5/172-(-8/172)$. Ten of the 21 recoveries lack baseline support and eleven already have it. This is descriptive accounting rather than a separation into independent causal effects (Appendix~\ref{app:posthoc}). Figure~\ref{fig:bank_selection} illustrates structural outcomes and compares selected joint success across coordinate banks and selection rules.

\subsection{Joint improvements coexist with physical-quality losses}
\label{sec:safety}
Among frozen-confidence-selected outputs, the additional PoseBusters-derived physical-quality checklist passes on 140/172 baseline predictions and 117/172 historical-guided predictions. Predictions passing both joint success and the checklist increase from 33 to 54: +12.21 points [5.70, 20.67] (Table~\ref{tab:physical_checklist_followup}). Across all candidates, checklist compliance decreases from 59.49\% to 37.76\%. Protein-overlap compliance declines, meaning more excessive overlap with the receptor, while the aggregate internal-quality rate changes little (Appendix~\ref{app:followup_pb}).

Historical guidance also increases candidate fatal clashes from 16.83\% to 27.22\%. Within 69 baseline ligand-correct/joint-failed observations, it recovers 26 joint successes but loses ligand accuracy in 14. At Top-40, historical and composition-matched guidance succeed in 96/172 and 106/172 banks, despite the higher historical Top-1 count. Of 33 historical Top-1 improvements, 25 already have a successful baseline candidate. Detailed physical and ranked-prefix results appear in Appendix~\ref{app:followup_pb}.

\begin{table}[H]
\centering
\caption{Joint efficacy and the additional physical checklist on SurfDock. The first three rows count confidence-selected outputs out of 172; the last row uses 6,880 candidates per condition. Differences are percentage points with common group-bootstrap 95\% intervals.}
\label{tab:physical_checklist_followup}
\fitbox{\begin{tabularx}{\linewidth}{@{}Yrrr@{}}\toprule
Outcome & No guidance & Historical & Difference [95\% interval] \\\midrule
Joint success & 37 & 63 & $+15.12\ [8.00,24.63]$ \\
Physical checklist & 140 & 117 & $-13.37\ [-21.83,-6.52]$ \\
Joint and checklist & 33 & 54 & $+12.21\ [5.70,20.67]$ \\
Candidate checklist (\%) & 59.49 & 37.76 & $-21.73\ [-26.87,-15.31]$ \\
\bottomrule\end{tabularx}}
\end{table}

\subsection{Timing of donor correction}
\label{sec:timing}
With the same historical set and maximum correction budget, interleaved and terminal-only correction yield 63/172 and 65/172 joint successes. Their difference is $-1.16$ points [$-3.19$, 1.15], below the prespecified +3-point superiority target. Count-only guidance similarly yields 56/172 and 59/172. Terminal-only historical correction has fewer selected clashes (18 versus 22) but more candidate clashes (30.99\% versus 27.22\%). These results do not establish a need to intervene during diffusion rather than after sampling. We retain the reference schedule and report the timing and output-quality trade-off in Appendix~\ref{app:followup_timing}.
\FloatBarrier

\section{Discussion}
Both historical and trajectory-local policies improve selected joint success on the evaluated backbones. Much of the aggregate gain is recovered using historical counts and composition. Fixed historical membership has no resolved incremental advantage, and exact assignments can still fail at the pose level.

A prospective application is pose-hypothesis generation during analogue optimization. A co-complex of a related compound could supply a candidate donor assignment, while a frozen docking model explores how the modified ligand fits the pocket. CICG provides an interface for incorporating this assignment into pose generation. Useful deployment would require reliable atom mapping, evidence supporting the assumed coordination mode, and physical-quality assessment of the resulting poses.

The post-hoc fixed-selector analysis supports additional joint utility from changing the bank under the tested rule. Increased candidate support and fewer selection misses describe the gain.

More selected poses satisfy both joint success and the checklist, while checklist compliance deteriorates among selected outputs and across the bank. Radial correction can compromise ligand placement or protein compatibility. Joint success excludes fatal clashes, but its distance screen does not guarantee protein-overlap compliance. For the same unmodified and historical-guided banks, the observed effect on joint success changes sign between the overlap-based and radial/clash selectors. Both selected outputs and complete banks require physical-quality evaluation.

The reporting-stage input restriction changes the estimand and error composition: historical sets are exact for 73/100 collected and 69/86 assessable queries, excluding four exact and ten inexact sets. Appendix~\ref{app:fullcohort} retains the original cohort results. Initial-roster coverage (62/304) cannot be combined with the pooled effect to estimate full-population utility. Existing matching and consensus rules establish eligibility, not calibrated transfer reliability. A stricter historical-only screen covered no current query (Appendix~\ref{app:historical_screen}); a useful reliability-based abstention rule remains unvalidated.

Both backbones use the same queries, so their results are technical replication rather than independent biological validation. Only three magnesium queries remain assessable, alongside 22 zinc queries. Group identifiers and chronological reference restrictions do not eliminate protein-family dependence or pretrained-model exposure. The two fixed seeds provide limited sampling replication. Reference/site-linked sensitivities retain their particular contrasts. Broader validation requires a family/site-separated cohort, with eligibility, assessability, and analysis fixed before docking outcomes are observed.

Water occupancy, protonation, receptor rearrangement, and electronic states remain unresolved. Comparisons with generic guidance or reference-coordinate transfer under equal information and compute are still needed to attribute advantages to the correction algorithm itself. The MetalloDock operating path avoids native donor labels and ligand coordinates as model inputs, but uses holo-derived pocket locations and differs in preprocessing, refinement, and candidate budget. Stored physical-check flags are available for replay; the exact historical PoseBusters release and default thresholds were not pinned.

\section{Conclusion}
CICG enables supplied-donor correction of frozen diffusion dockers without retraining. Historical and trajectory-local policies improve joint success on the evaluated domain, while the incremental benefit of historical membership remains unresolved. Coordinate guidance adds utility under the tested donor-aware selector, with losses in ligand accuracy and bank-wide protein-overlap compliance. The timing study does not establish diffusion-time intervention as necessary. Reliable transfer/abstention, physical-quality preservation, and independent family/site validation remain prerequisites for broader claims.

\label{page:mainend}
\clearpage

\section*{Acknowledgments}
This work was partly supported by the KHIDI grant funded by the Korean government (MOHW) [No.RS-2025-02307233, No.RS-2026-25613012], the NRF or IITP grants funded by the Korean government (MSIT) [No.05-26-04-0094, No.RS-2026-25472075, No.RS-2025-02305581, No.RS-2025-25442338, and No.RS-2021-II211343], the ITIP grant funded by the Korean government (MOTIR) [No.RS-2026-25549946], the Research grant from SNU, and the Strategic Hub grant for International Research Collaboration of SNU.

Kyungsu Kim is affiliated with the School of Transdisciplinary Innovations, Seoul National University; Department of Biomedical Science, Seoul National University; Interdisciplinary Program in Artificial Intelligence (IPAI), Seoul National University; Graduate School of AI, Seoul National University; Medical Research Center, Seoul National University; and AI Institute, Seoul National University.

\bibliography{main}

@inproceedings{corso2023diffdock,
  title = {{DiffDock}: Diffusion Steps, Twists, and Turns for Molecular Docking},
  author = {Corso, Gabriele and St{\"a}rk, Hannes and Jing, Bowen and Barzilay, Regina and Jaakkola, Tommi},
  booktitle = {International Conference on Learning Representations},
  year = {2023},
  url = {https://arxiv.org/abs/2210.01776}
}

@article{cao2025surfdock,
  title = {{SurfDock} is a surface-informed diffusion generative model for reliable and accurate protein--ligand complex prediction},
  author = {Cao, Duanhua and Chen, Mingan and Zhang, Runze and Wang, Zhaokun and Huang, Manlin and Yu, Jie and Jiang, Xinyu and Fan, Zhehuan and Zhang, Wei and Zhou, Hao and Li, Xutong and Fu, Zunyun and Zhang, Sulin and Zheng, Mingyue},
  journal = {Nature Methods},
  volume = {22},
  number = {2},
  pages = {310--322},
  year = {2025},
  doi = {10.1038/s41592-024-02516-y}
}

@article{buttenschoen2024posebusters,
  title = {{PoseBusters}: {AI}-based docking methods fail to generate physically valid poses or generalise to novel sequences},
  author = {Buttenschoen, Martin and Morris, Garrett M. and Deane, Charlotte M.},
  journal = {Chemical Science},
  volume = {15},
  number = {9},
  pages = {3130--3139},
  year = {2024},
  doi = {10.1039/D3SC04185A}
}

@inproceedings{bansal2024universal,
  title={Universal Guidance for Diffusion Models},
  author={Bansal, Arpit and Chu, Hong-Min and Schwarzschild, Avi and Sengupta, Roni and Goldblum, Micah and Geiping, Jonas and Goldstein, Tom},
  booktitle={International Conference on Learning Representations},
  year={2024},
  url={https://openreview.net/forum?id=pzpWBbnwiJ}
}

@inproceedings{christopher2024pdm,
  title={Constrained Synthesis with Projected Diffusion Models},
  author={Christopher, Jacob K. and Baek, Stephen and Fioretto, Ferdinando},
  booktitle={Advances in Neural Information Processing Systems},
  volume={37},
  year={2024},
  url={https://proceedings.neurips.cc/paper_files/paper/2024/hash/a28af221f2f70be183afc16797a56b91-Abstract-Conference.html}
}

@article{zhang2026metallodock,
  title = {{MetalloDock}: Decoding Metalloprotein--Ligand Interactions via Physics-Aware Deep Learning for Metalloprotein Drug Discovery},
  author = {Zhang, Hui and Zhang, Xujun and Su, Qun and Zheng, Yangyang and Jiang, Linlong and Zhu, Kai and Gou, Qiaolin and Zhang, Odin and Li, Shi and Peng, Bo and Ni, Shaokai and Du, Yushen and Tang, Jiayi and Kang, Yu and Hsieh, Chang-Yu and Li, Dan and Chen, Wenteng and Hou, Tingjun and Pan, Peichen},
  journal = {Journal of the American Chemical Society},
  volume = {148},
  number = {3},
  pages = {3086--3101},
  year = {2026},
  doi = {10.1021/jacs.5c15876}
}

@article{santosmartins2014autodockzn,
 author = {Santos-Martins, Diogo and Forli, Stefano and Ramos, Maria Jo{\~a}o and Olson, Arthur J.},
 title = {{AutoDock4Zn}: An Improved {AutoDock} Force Field for Small-Molecule Docking to Zinc Metalloproteins},
 journal = {Journal of Chemical Information and Modeling},
 volume = {54}, number = {8}, pages = {2371--2379}, year = {2014},
 doi = {10.1021/ci500209e}
}

@article{wang2020gmdockzn,
 author = {Wang, Kai and Lyu, Nan and Diao, Hongjuan and Jin, Shujuan and Zeng, Tao and Zhou, Yaoqi and Wu, Ruibo},
 title = {{GM-DockZn}: A geometry matching-based docking algorithm for zinc proteins},
 journal = {Bioinformatics}, volume = {36}, number = {13},
 pages = {4004--4011}, year = {2020}, doi = {10.1093/bioinformatics/btaa292}
}

@article{minan2025geodirdock,
  title={Informed protein--ligand docking via geodesic guidance in translational, rotational and torsional spaces},
  author={Mi{\~n}{\'a}n, Ra{\'u}l and Gallardo, Javier and Ciudad, {\'A}lvaro and Molina, Alexis},
  journal={Nature Machine Intelligence}, volume={7}, pages={1555--1560}, year={2025},
  doi={10.1038/s42256-025-01091-x}
}

@article{elmachtoub2022spo,
 title={Smart ``Predict, then Optimize''},
 author={Elmachtoub, Adam N. and Grigas, Paul},
 journal={Management Science}, volume={68}, number={1}, pages={9--26},
 year={2022}, doi={10.1287/mnsc.2020.3922}
}

@inproceedings{mandi2022decision,
 title={Decision-Focused Learning: Through the Lens of Learning to Rank},
 author={Mandi, Jayanta and Bucarey, V{\'i}ctor and Tchomba, Maxime Mulamba Ke and Guns, Tias},
 booktitle={Proceedings of the 39th International Conference on Machine Learning},
 series={Proceedings of Machine Learning Research}, volume={162},
 pages={14935--14947}, year={2022},
 url={https://proceedings.mlr.press/v162/mandi22a.html}
}

@article{bates2021risk,
 title={Distribution-Free, Risk-Controlling Prediction Sets},
 author={Bates, Stephen and Angelopoulos, Anastasios and Lei, Lihua and Malik, Jitendra and Jordan, Michael I.},
 journal={Journal of the ACM}, volume={68}, number={6}, pages={43:1--43:34},
 year={2021}, doi={10.1145/3478535}
}

@article{angelopoulos2025ltt,
 title={Learn then Test: Calibrating Predictive Algorithms to Achieve Risk Control},
 author={Angelopoulos, Anastasios N. and Bates, Stephen and Cand{\`e}s, Emmanuel J. and Jordan, Michael I. and Lei, Lihua},
 journal={The Annals of Applied Statistics}, volume={19}, number={2}, pages={1641--1662}, year={2025},
 doi={10.1214/24-AOAS1998}
}

@inproceedings{prokhorenkova2018catboost,
 author={Prokhorenkova, Liudmila and Gusev, Gleb and Vorobev, Aleksandr and Dorogush, Anna Veronika and Gulin, Andrey},
 title={{CatBoost}: Unbiased Boosting with Categorical Features},
 booktitle={Advances in Neural Information Processing Systems},volume={31},year={2018},
 url={https://papers.neurips.cc/paper_files/paper/2018/hash/14491b756b3a51daac41c24863285549-Abstract.html}
}

@article{gasteiger1980,
 author={Gasteiger, Johann and Marsili, Mario},
 title={Iterative partial equalization of orbital electronegativity---a rapid access to atomic charges},
 journal={Tetrahedron},volume={36},number={22},pages={3219--3228},year={1980},
 doi={10.1016/0040-4020(80)80168-2}
}

@misc{rdkit2023,
  author    = {Landrum, Greg and Tosco, Paolo and Kelley, Brian and others},
  title     = {{rdkit/rdkit: 2023\_03\_1 (Q1 2023) Release}},
  year      = {2023},
  publisher = {Zenodo},
  doi       = {10.5281/zenodo.7880616},
  url       = {https://zenodo.org/records/7880616}
}

@article{bonardi2020tails,
 author={Bonardi, Alessandro and Nocentini, Alessio and Bua, Silvia and Combs, Jacob and Lomelino, Carrie and Andring, Jacob and Lucarini, Laura and Sgambellone, Silvia and Masini, Emanuela and McKenna, Robert and Gratteri, Paola and Supuran, Claudiu T.},
 title={Sulfonamide Inhibitors of Human Carbonic Anhydrases Designed through a Three-Tails Approach: Improving Ligand/Isoform Matching and Selectivity of Action},
 journal={Journal of Medicinal Chemistry},volume={63},number={13},pages={7422--7444},year={2020},doi={10.1021/acs.jmedchem.0c00733}
}

@article{errington2025interaction,
  author = {Errington, David and Schneider, Constantin and Bouysset, C{\'e}dric and Dreyer, Fr{\'e}d{\'e}ric A.},
  title = {Assessing interaction recovery of predicted protein-ligand poses},
  journal = {Journal of Cheminformatics},
  volume = {17},
  number = {1},
  pages = {76},
  year = {2025},
  doi = {10.1186/s13321-025-01011-6}
}

@misc{rdkit2026analysis,
  author    = {Landrum, Greg and Tosco, Paolo and Kelley, Brian and others},
  title     = {{rdkit/rdkit: 2025\_09\_6 (Q3 2025) Release}},
  year      = {2026},
  publisher = {Zenodo},
  doi       = {10.5281/zenodo.18797641},
  url       = {https://zenodo.org/records/18797641}
}

@article{kabsch1976,
 author={Kabsch, Wolfgang},
 title={A solution for the best rotation to relate two sets of vectors},
 journal={Acta Crystallographica Section A},volume={32},number={5},pages={922--923},year={1976},
 doi={10.1107/S0567739476001873}
}

@article{berman2000pdb,
 author={Berman, Helen M. and Westbrook, John and Feng, Zukang and Gilliland, Gary and Bhat, T. N. and Weissig, Helge and Shindyalov, Ilya N. and Bourne, Philip E.},
 title={The {Protein Data Bank}},
 journal={Nucleic Acids Research},volume={28},number={1},pages={235--242},year={2000},doi={10.1093/nar/28.1.235}
}

@inproceedings{corso2024diffdockl,
  title={Deep Confident Steps to New Pockets: Strategies for Docking Generalization},
  author={Corso, Gabriele and Deng, Arthur and Polizzi, Nicholas and Barzilay, Regina and Jaakkola, Tommi},
  booktitle={International Conference on Learning Representations},
  year={2024},
  url={https://proceedings.iclr.cc/paper_files/paper/2024/hash/db334db287337b2a365120b524300ef3-Abstract-Conference.html}
}

@article{clemente2024mbd,
  author = {Clemente, Camila M. and Prieto, Juan M. and Mart{\'i}, Marcelo},
  title = {Unlocking Precision Docking for Metalloproteins},
  journal = {Journal of Chemical Information and Modeling},
  year = {2024},
  volume = {64},
  number = {5},
  pages = {1581--1592},
  doi = {10.1021/acs.jcim.3c01853}
}

@article{lai2026forcefield,
 author = {Lai, Helen and Wang, Tingyu and Sirelkhatim, Hassan and Eaton, Joe and Huang, Howard and Rees, Brad and Engkvist, Ola and Janet, Jon Paul and Wang, Xiaoyun and Tibo, Alessandro},
 title = {Improving protein-ligand complex generation with force field guidance},
 journal = {Journal of Cheminformatics},
 volume = {18},
 pages = {55},
 year = {2026},
 doi = {10.1186/s13321-026-01198-2}
}

@inproceedings{nguyen2026stein,
 author = {Nguyen, Van Khoa and Blond{\'e}, Lionel and Kalousis, Alexandros},
 title = {Stein Diffusion Guidance: Training-Free Posterior Correction for Sampling Beyond High-Density Regions},
 booktitle = {Proceedings of the 43rd International Conference on Machine Learning},
 series = {Proceedings of Machine Learning Research},
 volume = {306},
 pages = {123434--123464},
 year = {2026},
 url = {https://proceedings.mlr.press/v306/nguyen26v.html}
}

@article{lee2016cryptic,
  author = {Lee, Chul-Jin and Liang, Xiaofei and Wu, Qinglin
            and Najeeb, Javaria and Zhao, Jinshi
            and Gopalaswamy, Ramesh and Titecat, Marie
            and Sebbane, Florent and Lemaitre, Nadine
            and Toone, Eric J. and Zhou, Pei},
  title = {Drug design from the cryptic inhibitor envelope},
  journal = {Nature Communications},
  volume = {7},
  pages = {10638},
  year = {2016},
  doi = {10.1038/ncomms10638}
}
\bibliographystyle{plainnat}
\clearpage
\appendix
\section{Implementation and evaluation details}
\label{app:implementation}
\subsection{Correction parameters}
The correction operates in the pretrained sampler's native translation, rotation-vector, and torsion coordinates. At each scheduled hook, it starts from the current sampled pose, computes the donor-distance residual and Jacobian, and accepts a backtracked update only when the residual decreases and the per-update Cartesian displacement satisfies the cap. Translations and atom positions are represented in \AA, and rotation-vector and torsion components in radians. The finite-difference steps and metric weights refer to these coordinates. The numerical settings are listed in Table~\ref{tab:operator}; the metal--element targets are given in Table~\ref{tab:targets}.

\begin{table}[!htbp]
\centering
\caption{Numerical settings of the reference SurfDock correction. The finite-difference step and metric refer to the stored kinematic coordinates.}
\label{tab:operator}
\fitbox{\begin{tabularx}{\linewidth}{@{}P{0.44\linewidth}Y@{}}\toprule
Quantity&Value\\\midrule
SurfDock reverse steps&20\\
Activated steps&After 18 and 19, zero-based indexing\\
Samples per query and seed&40\\
Coordinates&Translation, rotation vector, rotatable-bond torsions\\
Residual scale&$0.10\ang$\\
Jacobian&Central finite differences with step $10^{-5}$\\
Metric diagonal&1 for translation/rotation; 8 for torsions\\
Linear solve&Metric-weighted SVD pseudoinverse\\
Relative singular-value cutoff&$10^{-10}$\\
Relinearizations per intervention&At most 5\\
Stopping tolerance&$\|r\|_2\le10^{-3}$\\
Backtracking scales&$\{2^{-b}:b=0,\ldots,15\}$\\
Per-update heavy-atom RMS limit&$0.25\ang$\\
Acceptance&Finite coordinates, reduced residual norm, displacement limit\\
Clash and stereochemistry guard&None inside the correction\\
Angle, exclusion, or strain terms&None\\\bottomrule
\end{tabularx}}
\end{table}
\begin{table}[!htbp]
\caption{Fixed metal--donor-element target distances $d^{\rm tar}_{s_m,e}$ in \AA. Dashes indicate unsupported pairs. These empirical targets are shared with the earlier seven-hook study; the correction algorithms and schedules differ.}
\label{tab:targets}
\centering
\fitbox{\begin{tabular}{@{}lrrr@{}}\toprule
Metal & N & O & S\\\midrule
Ca &---&2.40&---\\
Co &2.10&2.05&2.25\\
Fe &2.10&2.05&2.30\\
Mg &2.20&2.10&---\\
Mn &2.20&2.15&2.35\\
Zn &2.10&2.05&2.30\\\bottomrule
\end{tabular}}
\end{table}

Equation~\eqref{eq:update} follows by writing $z=M^{1/2}\Delta q$. The linearized least-squares problem is $\min_z\|J M^{-1/2}z+r\|_2$. The minimum-Euclidean-norm solution in $z$ is $-(JM^{-1/2})^+r$, yielding the minimum-$M$-norm correction in $\Delta q$. The rank cutoff and nonlinear line search define a local numerical correction. The displacement restriction is a Cartesian RMS cap on each accepted relinearization.

\paragraph{Kinematic mapping.}
Each correction call uses the sampled ligand as its reference conformer. Rigid rotation is about that conformer's centroid, followed by translation. Rotatable edges and their moving-atom masks are processed in the stored graph order; the first endpoint is stationary and the second belongs to the moving component. The flexible proposal is then Kabsch-aligned to the rigid proposal \citep{kabsch1976}. Retraction adds translations, left-composes rotation matrices, and wraps accumulated torsions. The metric weights are one for each translation and rotation component and eight for each torsion. The Cartesian displacement test permits a numerical tolerance of $10^{-12}\ang$. No collision or stereochemical test enters the line-search acceptance rule.

\subsection{Historical data and transfer rules}
\label{app:historical}
The reference collection retains 2,344 usable records from 2,423 raw records. References are released before 1 January 2023 and strictly before the corresponding query. Candidates must have the same metal and a different ligand component and canonical isomeric molecular graph. The original 304-query roster combines 179 non-overlapping entries from a 400-complex collection with 125 entries from a separate post-2023 collection. Of these, 112 have a reference at the matched target/site and 62 receive a donor set. Native correctness is evaluated after transfer and does not select the replay population.

\paragraph{Receptor-site compatibility.}
A first shell consists of every standard protein residue with a non-hydrogen/deuterium atom within $3.0\ang$ of the metal. At least one reference/query chain pair must have global-alignment sequence identity and coverage in both directions of at least 90\%. Alignment scores are one for a match, zero for a mismatch, $-1$ for gap opening, and $-0.1$ for extension. The first optimal alignment is used. The complete sets of aligned first-shell residue positions and residue identities must agree. The $3.0\ang$ threshold defines protein first-shell membership.

\paragraph{Atom transfer.}
Global maximum-common-substructure matching preserves elements, formal charges, exact bond orders, ring membership, and complete rings. It uses a five-second timeout and requires at least six atoms and 50\% of the smaller ligand. At most 2,048 matches per ligand are enumerated. All reference donors must be included. Donor-local signatures compare element, formal charge, aromaticity, degree, and neighboring atom/bond states. A query receives a set only when the accepted projections agree.

The donor-centered extension instead matches atoms within two graph edges of the reference donors together with shortest paths connecting donor pairs. It uses an atom-mapped substructure query, ring/bond consistency, the same donor-local checks, and at most 4,096 matches. Multiple distinct projected donor sets are rejected rather than merged by evaluator symmetry. These transfer match calls do not explicitly enable chirality-sensitive matching, and no extra protonation or tautomer normalization is applied. This differs from the chirality-preserving whole-graph symmetry used in evaluation and in the selection-only comparator.

\paragraph{Reference consensus.}
Repeated references with the same canonical isomeric ligand graph contribute one vote only if they map successfully and agree. The graph-deduplicated winner must be unique, receive at least two graph votes and at least two-thirds of the votes, and remain the unique winner after any single graph vote is removed. The minimum-vote and vote-share thresholds are not reapplied inside this leave-one-vote-out check. Existing strict-rule assignments are retained; donor-centered assignments precede consensus assignments. Ties and nonunique mappings lead to abstention.

\begin{table}[!htbp]
\caption{Historical donor transfer on the original 304-query roster. Coverage is relative to all queries; precision is relative to supplied sets. Extension rows contain disjoint additions.}
\label{tab:historical}
\centering
\fitbox{\begin{tabularx}{\linewidth}{@{}Yrrrr@{}}\toprule
Transfer rule&Supplied&Correct&Coverage (\%)&Precision (\%)\\\midrule
Strict site and global-graph matching&33&25&10.86&75.76\\
Donor-centered core additions&13&9&---&69.23\\
Graph-deduplicated consensus additions&16&11&---&68.75\\
Combined transfer&62&45&20.39&72.58\\\bottomrule
\end{tabularx}}
\end{table}

The donor-centered extension supplies 13 of 61 previously unsupported queries. Consensus supplies 16 of 18 disagreement cases, all in one group. Combined transfer precision is 72.58\% (17-group bootstrap interval [57.89, 84.75]\%); 12 of its 17 errors retain the correct donor count. Reference donor names are mapped to indices only when the archived chemical-component heavy-atom name/element order matches the ligand graph. The supplement contains the execution contracts, candidate reference pairs, and reference metadata.

\paragraph{Additional-query selection.}
The additional cohort was selected from two previously screened structural collections: records through 15 July 2026 and a recent-release collection spanning 22 July--23 September 2026. Eligible records had to receive a historical contract, avoid Protein Data Bank entries already present in the initial cohort, and pass the recorded technical preparation criteria. An initial 60-query candidate set used square-root allocation by applicable metal counts, prioritizing recent releases and then a fixed hash order within metal. From this ordered set, fixed quotas selected 38 queries: three calcium, four cobalt, one iron, eight magnesium, one manganese, and 21 zinc complexes. Selection used neither native query-donor labels nor docking outcomes. These quotas deliberately broaden metal representation; they do not define a random sample of all metalloproteins.

\begin{table}[!htbp]
\centering
\caption{Metal composition of the evaluated historical-transfer cohorts. The combined sample includes the initial and additional queries.}
\label{tab:metal_composition}
\fitbox{\begin{tabularx}{\linewidth}{@{}Yrrr@{}}\toprule
Metal&Initial (62)&Additional (38)&Combined (100)\\\midrule
Calcium&14&3&17\\
Cobalt&21&4&25\\
Iron&9&1&10\\
Magnesium&2&8&10\\
Manganese&15&1&16\\
Zinc&1&21&22\\\bottomrule
\end{tabularx}}
\end{table}

The initial 62 and additional 38 queries have disjoint Protein Data Bank entries and no shared frozen group identifiers. Group construction uses receptor/site and reference connectivity for the strict-rule subset and sets of supporting reference identifiers for the extension and additional subsets. The resulting 17 and 22 groups are the dependence units retained in the analyses; distinct group identifiers do not establish protein-family independence. The combined 100-query sample has 39 groups, with 73 correct and 27 incorrect historical sets. The pooled follow-up applies the input-domain restriction described in Section~\ref{sec:evaluation}. Coverage of 62/304 applies only to the original screened roster.

\begin{table}[!htbp]
\centering
\caption{Input-domain filtering by metal. Each entry is collected queries / assessable queries. Exclusion reflects the stereochemistry checker's supported representation, not chemical invalidity or generated-pose quality.}
\label{tab:domainmetals}
\fitbox{\begin{tabularx}{\linewidth}{@{}Yrrr@{}}\toprule
Metal&Initial&Additional&Combined\\\midrule
Calcium&14 / 8&3 / 2&17 / 10\\
Cobalt&21 / 21&4 / 4&25 / 25\\
Iron&9 / 9&1 / 1&10 / 10\\
Magnesium&2 / 1&8 / 2&10 / 3\\
Manganese&15 / 15&1 / 1&16 / 16\\
Zinc&1 / 1&21 / 21&22 / 22\\\midrule
Total&62 / 55&38 / 31&100 / 86\\\bottomrule
\end{tabularx}}\end{table}

\subsection{Pose-level success and stereochemistry}
\label{app:evaluator}
Native ligand coordinates $\mathbf x_i^\star$ and donor annotations $D^\star$ are fixed evaluation inputs in the receptor frame, separate from the historical-selector inputs. The archived annotations are mapped by ligand atom identity; the evaluation does not infer $D^\star$ from generated poses. Query labels and their atom-name/index correspondence are included in \texttt{evaluation/references.json} and the mapping provenance ledger. The release supports evaluation conditional on these annotations, but does not reconstruct the upstream chemical curation of every native donor label. This limits independent auditing of annotation ambiguity. With $K_\star=|D^\star|$, let $N_{K_\star}(y)$ be the nearest $K_\star$ ligand heavy atoms to the metal in pose $y$, including all heavy-atom elements and resolving distance ties by atom index. A coordination witness is a chirality-preserving whole-ligand mapping $\pi$, where $\pi(i)$ identifies the predicted-ligand atom corresponding to native atom $i$.
The witness must satisfy
\begin{align}
 N_{K_\star}(y)&=\pi(D^\star),\label{eq:membership}\\
 \max_{i\in D^\star}\left|\|\mathbf x_{\pi(i)}(y)-\mathbf z_m\|_2-
                          \|\mathbf x_i^\star-\mathbf z_m\|_2\right|&\le0.50\ang,\label{eq:distance_eval}\\
 \max_{i\in D^\star}\angle\big(\mathbf x_{\pi(i)}(y)-\mathbf z_m,
                          \mathbf x_i^\star-\mathbf z_m\big)&\le30^\circ.\label{eq:angle_eval}
\end{align}
All three conditions must hold under one mapping. The
evaluator's allowed mappings are defined separately from
the input-graph automorphisms $\mathcal G_L$ used for reranking in Equation~\eqref{eq:scoring}. Chemically cleaned, chirality-preserving full-graph mappings are enumerated with a cap of 10,000; cap exceedance fails the evaluation rather than truncating its witness search. Whole-ligand heavy-atom RMSD is minimized separately over the allowed mappings, without spatial alignment, and must be strictly below $2.0\ang$. Its minimizing mapping need not be the coordination witness.

Stereochemistry is checked against the chemically specified input ligand. Each supported carbon tetrahedral center must retain the sign of its normalized signed triple product, whose magnitude is at least 0.02. For each specified E/Z double bond, let $\tau_{\rm in}$ and $\tau_{\rm out}$ be the dihedral angles in radians for the same ordered atom tuple $(i_1,i_2,i_3,i_4)$ in the chemically specified input and predicted conformers. The outer atoms $(i_1,i_4)$ are the stereochemical atoms returned by \texttt{GetStereoAtoms()}, while $(i_2,i_3)$
are the bond's begin and end atom indices. Both conformers require $|\sin\tau|\le0.5$, and $\cos\tau_{\rm in}$ and $\cos\tau_{\rm out}$ must have
matching signs. Specified non-carbon tetrahedral centers and ligand-internal metals are outside the checker's supported scope. The original full-cohort analyses count these inputs as failures. Joint-success tables in the main text pool the input-domain subset: 55 of 62 initial queries and 31 of 38 additional queries, totaling 86. This reporting-stage restriction retains every assessable query, regardless of its predicted validity or donor-set correctness. The 14 excluded input graphs and exclusion reasons are listed in \texttt{assessability/INPUT\_\allowbreak DOMAIN\_\allowbreak LEDGER.tsv}.

The geometry-valid flag is the conjunction of finite coordinates, successful RDKit sanitization \citep{rdkit2026analysis} of the input graph with predicted coordinates, preserved supported stereochemistry, and no fatal receptor clash. A fatal clash is any ligand/receptor heavy-atom separation strictly below $1.2\ang$; nonfinite ligand coordinates also fail. The receptor coordinate reader retains non-hydrogen ATOM/HETATM records in the prepared receptor. Evaluation exceptions are recorded as failures, including a failed sanitization flag. These tests are narrower than the full PoseBusters battery \citep{buttenschoen2024posebusters}. Joint success requires ligand success, a common coordination witness, and geometry validity.

Separate identity, radial, and directional diagnostics refer to the same selected pose. Each marginal tests existence over complete assignments, but their witnesses can differ; multiplying these marginal indicators would not reconstruct the joint criterion.

\subsection{Paired statistical analysis}
\label{app:statistics}
Query-level effects average the two fixed-seed paired differences before resampling. For each of 20,000 replicates, the bootstrap samples $G$ of the $G$ frozen group identifiers uniformly with replacement, concatenates all queries in each sampled group with their multiplicities, and averages the query effects. Groups are sampling units, not equally weighted group means. All conditions and both seeds stay paired. Two-sided intervals use the linearly interpolated 2.5th and 97.5th percentiles. Thus 124 or 200 query--seed outcomes do not constitute independent biological sample sizes.

\paragraph{Common pooled analysis.}
All full 86-query contrasts in the revised main text and Appendix~\ref{app:followups} use \texttt{AUDIT\_\allowbreak CANONICAL\_\allowbreak 86\_\allowbreak 33\_\allowbreak V1}. We average seeds 42 and 43 within each query, sample 33 original structural groups with replacement, and divide the sampled sum of query effects by the sampled query count. Groups with more queries consequently contribute more observations; this is not an equally weighted average of group means. The shared 20,000 draws use NumPy \texttt{default\_\allowbreak rng} (PCG64), seed 20263001. Their little-endian int32 hash and complete query/group ordering are recorded in \texttt{canonical/BOOTSTRAP\_\allowbreak MANIFEST.json}. The 2.5th and 97.5th percentiles use linear interpolation. Matrix and reference-loop implementations agree. The canonical replay contains 102 contrasts, including the full-cohort policy, selector, timing, second-backbone, and stored-checklist comparisons. Earlier subset analyses retain their own group populations and recorded intervals. For the pooled 86-query analyses, empirical counts and point estimates are unchanged from the submitted version; some confidence-limit endpoints differ slightly because the revised analysis applies a common 33-group resampling schedule instead of the originally submitted contrast-specific schedules.

The fixed-selector contrast reranks each bank independently with Equation~\eqref{eq:scoring}. Reusing the baseline-selected sample ID on the guided bank instead defines a different paired experiment (Appendix~\ref{app:followup_statistics}). Paired outcome keys and selection provenance are checked before calculating intervals.

\paragraph{Earlier cohort-specific analyses.}
The original analysis uses \texttt{numpy.random.default\_\allowbreak rng}. For the initial cohort, pooled historical-minus-baseline, historical-minus-nearest, nearest-minus-baseline, and native-minus-baseline comparisons use random states 20272001, 20272002, 20272003, and 20272004, respectively. The expanded analysis uses 20263001, 20263003, 20263002, and 20263004 for the same contrasts. The post-hoc comparisons share state 20272001 for the initial cohort and 20263001 for the additional and combined cohorts. Per-seed and correctness-stratum random states are recorded in \texttt{METHODS\_\allowbreak SETTINGS.json}.

Correctness strata are formed before calling the bootstrap function, and only groups represented in that stratum are sampled. All analyzed strata are nonempty. Contrast-specific random streams can produce slightly different Monte Carlo percentile endpoints for identical effect vectors; the original full-cohort intervals and streams are retained in Appendix~\ref{app:fullcohort}. The expanded historical-minus-nearest lower endpoint is exactly zero before rounding. The supplementary analysis reproduces the four central pooled historical-versus-baseline/nearest intervals from the stored query outcomes.

\paragraph{Assessable-domain analysis.}
For the assessable-domain analysis, an input-only mask removes specified non-carbon tetrahedral centers carrying a stereochemical label and ligand-internal metal atoms. It gives 55/15, 31/18, and 86/33 query/group counts for the initial, additional, and combined domains. We filter the query ledger before resampling, including for correctness strata. The earlier four-condition and selection-only contrasts within each cohort share 20,000 bootstrap draws (state 20272001 for the initial cohort and 20263001 for the additional and combined cohorts). Integer multiplicities accumulate the two-seed binary differences before division, retaining exact zero contrasts. Linear-interpolated 2.5th and 97.5th percentiles form the intervals. Intervals are recomputed after applying the input-domain mask. The script \texttt{code/analyze\_\allowbreak assessable.py} records the mask, counts, group numbers, and all numerical settings. Those intervals are retained with the original four-condition and selection-only assessable records. The initial composition-matched report used separate contrast-specific streams. Their settings and intervals remain in the historical record; the current pooled policy and follow-up comparisons use the common pooled schedule above.

Historical transfer precedes its native-label audit. The second seed and nearest-donor comparison are follow-ups to the first-seed experiment. The follow-up analysis criterion requires a positive historical-minus-baseline effect at both seeds and a positive pooled lower confidence limit. Correctness strata and Top-$k$ outcomes are secondary. The overlap-based reranking rule and group-deletion diagnostics were specified after the original outcomes were inspected and are post-hoc; their intervals are not multiplicity-adjusted.

\subsection{Data and implementation availability}
\label{app:data_implementation}
The code is maintained in a private GitHub repository at \url{https://github.com/pe7ace/CICG} and will be made publicly available upon publication of the peer-reviewed article.
The submitted supplement preserves the original four-condition 100-query analysis as 800 pose banks, 32,000 pose-level outcomes, query/group assignments, input and evaluation graphs, and historical-transfer provenance. Its original selection-only and bank-by-selector outputs remain available, along with the composition-matched files. The follow-up archive, \texttt{CICG\_\allowbreak FOLLOWUP\_\allowbreak REPRO\_\allowbreak ADDON.zip}, extends coordinate and execution provenance and is self-contained for the following CPU analyses. Extract it into a directory of the same name and run from that directory:
\begin{verbatim}
python verify_checksums.py
python canonical/replay.py --out replay/analysis
python replay_execution_audits.py --with-raw \
    --out replay/execution
python replay_selectors.py --out replay/selectors
python replay_raw_coordinates.py --study all --out replay/raw
\end{verbatim}
The canonical command reconstructs 102 pooled contrasts, 48 historical interval checks, and stored-checklist summaries. The execution audit verifies correction hooks, retained parent coordinates, fixed memberships, and matching inputs. Selector replay restricts input access to the ligand graph, historical information, candidate coordinates, and scoring metadata. Coordinate replay evaluates 27,520 DiffDock-L poses, 20,640 terminal-control poses, and 86 MetalloDock outputs without alignment, minimization, or stereochemical repair. These tiers do not run new sampling or the PoseBusters checker.

The follow-up analysis environment records Python 3.11, NumPy 2.4.6, and pandas 3.0.3. Coordinate and selector replay additionally uses SciPy 1.17.1 and RDKit 2025.09.6. Requirements files and file-level hashes are included. The original four-condition analysis and its earlier verification environment are retained in the submitted package rather than retroactively relabeled as the follow-up environment.

DiffDock-L raw SDF text, filenames, hashes, and saved parent arrays are held in lossless NPZ containers; terminal and MetalloDock outputs retain their recorded coordinates and atom order. Source snapshots, checkpoint identifiers, hook tables, and paired outcome ledgers connect these outputs to the described experiments. Model weights are not redistributed. Recovered checkpoint hashes identify files available during the audit, not runtime attestations of the historical loaded bytes. Contemporaneous container digests, some invocation arguments, and RNG/noise tensors were not retained. These omissions limit end-to-end sampling recreation but not the provided stored-output analyses.

For PoseBusters, all 27,520 stored rows and the 688 selected-pose records are included. The checklist columns, missing-value treatment, and receptor contents are specified in \texttt{POSEBUSTERS\_\allowbreak CONFIG.json}. Exact aggregation is reproducible; the historical package release and its default-threshold YAML were not bound to the executed environment. Candidate installations recovered later are distinguished from an attested original runtime. The execution and data limitations for each study are listed in \texttt{REPRODUCTION\_\allowbreak LEVELS.json} and the study manifests.

\FloatBarrier
\section{Additional results for historical guidance}
\label{app:additional}
\subsection{Original full-cohort analyses}
\label{app:fullcohort}
The original analyses use all 62 initial and 100 combined queries, counting inputs outside the stereochemistry checker's scope as joint failures. Table~\ref{tab:full_summary} reports the original full-cohort paired effects and contrast-specific confidence intervals. These are the original analysis populations; the pooled analysis in the main text uses the 86-query assessable domain. Restricting the joint endpoint to the supported input domain changes the estimand and its group set, but does not add successful predictions. The reporting-stage mask and reduced-domain outcomes are given in Appendix~\ref{app:implementation} and the main text.
\begin{table}[!htbp]
\centering
\caption{Original full-cohort analysis, with unsupported inputs counted as joint failures. Entries are paired two-seed effects in percentage points and the original 95\% intervals.}
\label{tab:full_summary}
\fitbox{\begin{tabularx}{\linewidth}{@{}Yrr@{}}\toprule
Comparison&Initial: 62 queries&Combined: 100 queries\\
&17 groups&39 groups\\\midrule
Historical $-$ no guidance&+12.10 [4.35, 23.17]&+13.00 [6.70, 21.25]\\
Nearest-$K$ $-$ no guidance&+10.48 [3.77, 19.49]&+9.50 [4.50, 15.28]\\
Native $-$ no guidance&+17.74 [7.89, 30.23]&+16.50 [9.04, 24.77]\\
Historical $-$ Nearest-$K$&+1.61 [$-2.08$, 8.57]&+3.50 [0.00, 8.79]\\\midrule
Donor reranking $-$ no guidance&+8.06 [1.88, 18.52]&+7.00 [2.23, 13.58]\\
Historical $-$ donor reranking&+4.03 [$-3.41$, 9.65]&+6.00 [1.10, 10.99]\\\bottomrule
\end{tabularx}}
\end{table}

\paragraph{Initial assessable cohort.}
Tables~\ref{tab:primary_counts}--\ref{tab:primary_effects} retain the 55-query version of the comparisons.
\begin{table}[!htbp]
\centering
\caption{Joint Top-1 success on the 55 assessable queries in 15 groups from the initial 62-query cohort. Each condition produces 40 poses at each seed; the composition-matched condition was evaluated in a separate follow-up. The input-domain restriction is a reporting-stage analysis; the original full-cohort results are retained in Appendix~\ref{app:fullcohort}.}
\label{tab:primary_counts}
\fitbox{\begin{tabularx}{\linewidth}{@{}Yrrr@{}}\toprule
Condition&Seed 42&Seed 43&Mean success (\%)\\\midrule
No guidance&10/55&13/55&20.91\\
Historical donors&19/55&19/55&34.55\\
Nearest-$K$ (count only)&16/55&20/55&32.73\\
Composition-matched Nearest-$K$&16/55&19/55&31.82\\
Native-donor oracle&21/55&24/55&40.91\\
\bottomrule\end{tabularx}}
\end{table}

\begin{table}[!htbp]
\centering
\caption{Paired joint Top-1 effects on 55 assessable initial queries. Effects average the two seeds within query and intervals resample 15 groups. The upper block uses common draws from the original assessable reanalysis; the lower block gives the separately specified composition-matched follow-up. Differences are percentage points.}
\label{tab:primary_effects}
\fitbox{\begin{tabularx}{\linewidth}{@{}Yrr@{}}\toprule
Comparison&Difference&95\% confidence interval\\\midrule
Historical donors $-$ no guidance&+13.64&[5.26, 26.67]\\
Nearest-$K$ (count only) $-$ no guidance&+11.82&[4.55, 22.13]\\
Native-donor oracle $-$ no guidance&+20.00&[9.62, 33.67]\\
Historical donors $-$ Nearest-$K$ (count only)&+1.82&[$-2.38$, 9.76]\\\midrule
Composition-matched Nearest-$K$ $-$ no guidance&+10.91&[4.00, 21.05]\\
Historical donors $-$ composition-matched Nearest-$K$&+2.73&[0.00, 10.17]\\
\bottomrule\end{tabularx}}
\end{table}

\subsection{Seed-specific paired comparisons}
Table~\ref{tab:seed_effects} gives the per-seed transitions and confidence intervals. An improvement is a query changing from failure to success relative to the stated comparator; a regression is the reverse. Summing these transitions across seeds counts query--seed outcomes, not distinct queries.
\begin{table}[!htbp]
\caption{Seed-specific Top-1 comparisons on 62 queries. Effects and group-cluster 95\% confidence intervals are percentage points.}
\label{tab:seed_effects}
\centering
\fitbox{\begin{tabularx}{\linewidth}{@{}cYrrr@{}}\toprule
Seed&Comparison&Difference&\shortstack{Improvements/\\regressions}&95\% interval\\\midrule
42&Historical $-$ no guidance&+14.52&10 / 1&[3.77, 30.36]\\
43&Historical $-$ no guidance&+9.68&11 / 5&[3.64, 18.37]\\
42&Nearest-$K$ $-$ no guidance&+9.68&7 / 1&[1.54, 20.29]\\
43&Nearest-$K$ $-$ no guidance&+11.29&9 / 2&[4.00, 20.45]\\
42&Native $-$ no guidance&+17.74&12 / 1&[5.36, 35.59]\\
43&Native $-$ no guidance&+17.74&13 / 2&[7.50, 27.12]\\
42&Historical $-$ Nearest-$K$&+4.84&3 / 0&[0.00, 17.65]\\
43&Historical $-$ Nearest-$K$&$-1.61$&4 / 5&[$-4.92$, 0.00]\\\bottomrule
\end{tabularx}}
\end{table}

\subsection{Ranked-prefix success}
\begin{table}[!htbp]
\caption{Joint success among the first $k$ confidence-ranked candidates in the original 62-query cohort. Each entry counts queries out of 62. All banks contain 40 poses; hence Success@40 equals availability for these conditions.}
\label{tab:topk}
\centering
\fitbox{\begin{tabularx}{\linewidth}{@{}cYrrrrr@{}}\toprule
Seed&Condition&Top-1&Top-2&Top-4&Top-8&Top-40\\\midrule
42&No guidance&10&12&20&24&27\\
42&Historical donors&19&24&28&30&30\\
42&Nearest-$K$&16&24&30&34&34\\
42&Native-donor oracle&21&28&31&33&33\\\midrule
43&No guidance&13&19&24&26&27\\
43&Historical donors&19&23&26&27&29\\
43&Nearest-$K$&20&25&29&32&34\\
43&Native-donor oracle&24&28&31&33&35\\\bottomrule
\end{tabularx}}
\end{table}

Historical guidance has 47/124 successful query--seed banks at Top-2, compared with 31/124 for the baseline and 49/124 for Nearest-$K$. At Top-40, the corresponding totals are 59/124, 54/124, and 68/124. Native-donor guidance also has 68/124 Top-40 successes, although the individual successful queries need not be the same as those of Nearest-$K$. These prefix counts describe candidate production and ranking jointly.

At seed 42, the ten historical Top-1 improvements admit two different decompositions. Five selected trajectories become successful under correction and five were already successful but rise in rank. Separately, three improvements occur on queries with no successful pose anywhere in the baseline bank, and seven occur where a baseline success already exists. The partitions overlap and therefore are not additive.

\paragraph{Assessable correctness strata.}
Table~\ref{tab:subgroups} gives the corresponding initial-domain analysis for the original four conditions. These intervals use common draws within each represented stratum, independently of the composition-matched follow-up below.
\begin{table}[!htbp]
\centering
\caption{Retrospective contract-correctness strata among the 55 assessable initial queries. Each entry is a paired two-seed effect in percentage points with a 95\% interval. Groups represented within each stratum are resampled; correctness is evaluated from native labels after transfer.}
\label{tab:subgroups}
\fitbox{\begin{tabularx}{\linewidth}{@{}Yrr@{}}\toprule
Comparison&Correct: 42 queries&Incorrect: 13 queries\\&11 groups&9 groups\\\midrule
Historical donors $-$ no guidance&+21.43 [9.68, 39.06]&$-11.54$ [$-19.23$, 0.00]\\
Nearest-$K$ $-$ no guidance&+14.29 [5.41, 27.27]&+3.85 [0.00, 7.14]\\
Native-donor oracle $-$ no guidance&+21.43 [9.68, 39.06]&+15.38 [0.00, 26.92]\\
Historical donors $-$ Nearest-$K$&+7.14 [$-2.56$, 20.37]&$-15.38$ [$-26.00$, 0.00]\\
\bottomrule\end{tabularx}}
\end{table}

\subsection{Incorrect contracts and trajectory-specific donor selection}
\begin{table}[!htbp]
\caption{Same-cardinality historical errors in the original 62-query cohort: 12 queries, two seeds. Entries are pooled Top-1 effects in percentage points. Only the native-versus-baseline confidence interval is reported; other entries are descriptive point estimates.}
\label{tab:samek}
\centering
\fitbox{\begin{tabularx}{\linewidth}{@{}Yr@{}}\toprule
Comparison&Difference\\\midrule
Historical $-$ no guidance&$-12.50$\\
Nearest-$K$ $-$ no guidance&+4.17\\
Native $-$ no guidance&+16.67 [5.00, 28.57]\\
Historical $-$ Nearest-$K$&$-16.67$\\\bottomrule
\end{tabularx}}
\end{table}

Although both methods failed at Top-1 in all 17 incorrect-contract queries at seed 42, historical guidance removed baseline Top-40 support in four same-cardinality-error queries. With both seeds, the incorrect-contract Top-1 effect is negative. Correctness therefore affects more than whether a donor count is satisfied.

Nearest-$K$ selects a native-exact set at step 18 in 548 trajectories at seed 42 and 564 at seed 43, out of 2,480 trajectories each. At each seed, 45 queries have at least one such trajectory. This trace is an observation about the generated candidates, not native information used by the control. The donor-count input remains the historical prediction, which can itself be wrong.

\subsection{Additional cohort and expanded exactness analysis}
\label{app:expanded}
The additional 38 queries share no Protein Data Bank entry or frozen group identifier with the initial 62. The additional-cohort counts in Table~\ref{tab:fresh} are obtained by subtracting the primary-cohort counts from the combined 100-query counts and agree with the per-query records. The post-hoc intervals resample the additional cohort's 22 groups.
\begin{table}[!htbp]
\caption{Top-1 outcomes on the 38 additional queries in 22 groups. Mean rates average the two seeds.}
\label{tab:fresh}
\centering
\fitbox{\begin{tabularx}{\linewidth}{@{}Yrrr@{}}\toprule
Condition&Seed 42&Seed 43&Mean success (\%)\\\midrule
No guidance&8/38&6/38&18.42\\
Historical donors&12/38&13/38&32.89\\
Nearest-$K$&10/38&10/38&26.32\\
Native-donor oracle&11/38&14/38&32.89\\\bottomrule
\end{tabularx}}
\end{table}

\begin{table}[!htbp]
\caption{Retrospective exactness strata in the expanded 100-query analysis. Effects average two seeds; intervals resample the groups represented within each correctness stratum (Appendix~\ref{app:statistics}). The exactness strata are not prospective eligibility rules.}
\label{tab:expanded_strata}
\centering
\fitbox{\begin{tabularx}{\linewidth}{@{}Yrr@{}}\toprule
Comparison&Correct sets ($n=73$)&Incorrect sets ($n=27$)\\\midrule
Historical $-$ no guidance&+20.55 [11.93, 31.63]&$-7.41$ [$-14.58$, 0.00]\\
Historical $-$ Nearest-$K$&+7.53 [1.33, 15.25]&$-7.41$ [$-17.31$, 2.78]\\\bottomrule
\end{tabularx}}
\end{table}

The added queries contain 28 correct and ten incorrect transferred sets, bringing the total from 45/62 to 73/100. The full expanded exact stratum has a positive historical-versus-nearest interval, whereas the original exact stratum's interval includes zero. The combined analysis adds observations but reuses the original cohort. Its conditional result does not replace the overall historical-versus-nearest comparison, whose interval still touches zero.

\subsection{Concentration across metals and groups}
\begin{table}[!htbp]
\caption{Net Top-1 improvements by metal in the original 62-query cohort, summed over the two fixed seeds. These are query--seed differences in success counts, not per-metal success rates.}
\label{tab:metals}
\centering
\fitbox{\begin{tabularx}{\linewidth}{@{}Yrr@{}}\toprule
Metal&Historical $-$ no guidance&Native $-$ no guidance\\\midrule
Calcium&0&0\\
Cobalt&+7&+11\\
Iron&+2&+2\\
Magnesium&0&0\\
Manganese&+6&+9\\
Zinc&0&0\\\midrule
Total&+15&+22\\\bottomrule
\end{tabularx}}
\end{table}

Seven of the 17 structural/site groups contribute a positive historical-versus-baseline effect, one a negative effect, and nine zero. Metal-specific denominators appear in Table~\ref{tab:metal_composition}; each query contributes two seed outcomes. Zero net change can include offsetting improvements and regressions. These counts therefore delimit the observed aggregate gain without establishing uniform benefit within any metal class.

\subsection{Physical outcomes by evaluation domain}
\label{app:fullphysical}
Table~\ref{tab:safety} gives the 55-query initial assessable domain, Table~\ref{tab:initial_physical62} gives the original 62-query cohort, and Table~\ref{tab:full_physical100} gives all 100 collected queries, including the 14 input graphs outside joint-evaluation scope. Ligand and donor-identity predicates can still be examined on these inputs, and receptor clashes are measured directly from coordinates. The geometry-valid flag itself includes the stereo checker and therefore is false on unsupported inputs. Full-cohort and assessable-domain percentages answer different questions and are not pooled.

\subsection{Post-hoc preservation of baseline candidates}
\label{app:union}
After inspecting the initial seed-42 results, we also considered a union of the 40 unmodified and 40 historically guided candidates, ranked by the original SurfDock confidence model. No new coordinates were generated. The diagnostic was defined after the primary result and uses twice as many stored candidates. Its success predicate was recorded as \emph{Robust Full}, whereas the primary four-condition experiment explicitly uses the clash-free joint criterion. The union predicate has not been verified as identical to the main joint predicate.

\begin{table}[H]
\centering
\caption{Physical outcomes on the 55 assessable initial queries. Entries give seed 42, then seed 43. Selected-pose counts use 55 queries per seed; all-pose percentages use 2,200 candidates. The corresponding 62-query results are in Table~\ref{tab:initial_physical62}; Table~\ref{tab:full_physical100} instead reports the combined 100-query population.}
\label{tab:safety}
\fitbox{\begin{tabularx}{\linewidth}{@{}Yrrrr@{}}\toprule
Condition&\shortstack{Ligand-only\\Top-1 ($n$)}&\shortstack{Clash at\\Top-1 ($n$)}&\shortstack{All-pose\\clash (\%)}&\shortstack{All-pose\\valid (\%)}\\\midrule
No guidance&34, 39&4, 2&16.86, 15.64&83.14, 84.36\\
Historical donors&33, 36&6, 4&27.45, 26.00&72.55, 74.00\\
Nearest-$K$&36, 39&7, 7&26.32, 26.09&73.68, 73.91\\
Native-donor oracle&32, 38&6, 4&27.09, 27.09&72.91, 72.91\\
\bottomrule\end{tabularx}}
\end{table}

\begin{table}[H]
\centering
\caption{Physical outcomes on all 62 initial queries, including unsupported stereochemical inputs. Entries give seed 42, then seed 43. Selected counts use 62 queries per seed; candidate fractions use 2,480 poses per seed.}
\label{tab:initial_physical62}
\fitbox{\begin{tabularx}{\linewidth}{@{}Yrrrr@{}}\toprule
Condition&\shortstack{Ligand\\Top-1}&\shortstack{Top-1\\clashes}&\shortstack{All-pose\\clash (\%)}&\shortstack{Geometry-\\valid (\%)}\\\midrule
No guidance&37, 41&5, 4&18.51, 16.53&73.75, 74.84\\
Historical donors&34, 37&8, 8&30.20, 28.75&64.35, 65.65\\
Nearest-$K$ (count only)&39, 41&10, 10&27.78, 27.34&65.36, 65.56\\
Native-donor oracle&33, 39&10, 7&29.80, 29.56&64.68, 64.68\\\bottomrule
\end{tabularx}}
\end{table}

\begin{table}[H]
\centering
\caption{Full collected 100-query physical outcomes. Entries give seed 42, then seed 43. All-pose percentages use 4,000 poses per condition per seed. The separate geometry-valid flag includes supported stereochemistry.}
\label{tab:full_physical100}
\fitbox{\begin{tabularx}{\linewidth}{@{}Yrrrrr@{}}\toprule
Condition&Ligand&Identity&\shortstack{Top-1\\clashes}&\shortstack{All-pose\\clash (\%)}&\shortstack{All-pose\\valid (\%)}\\\midrule
No guidance&62, 62&36, 40&9, 6&17.77, 16.52&71.12, 71.92\\
Historical donors&57, 59&58, 60&15, 15&29.60, 29.85&62.67, 62.50\\
Nearest-$K$&63, 64&51, 51&14, 13&26.00, 26.52&63.62, 63.10\\
Native-donor oracle&55, 61&68, 71&17, 13&29.07, 29.32&63.00, 62.27\\
\bottomrule\end{tabularx}}
\end{table}

\begin{table}[!htbp]
\caption{Seed-42 candidate-union diagnostic on 62 queries. Success entries retain the diagnostic's Robust Full definition. Success@40 is a ranked prefix of an 80-pose union, not its full-bank availability.}
\label{tab:union}
\centering
\fitbox{\begin{tabularx}{\linewidth}{@{}Yrrr@{}}\toprule
Outcome&\shortstack{No guidance\\(40 poses)}&\shortstack{Historical\\(40 poses)}&\shortstack{Union\\(80 poses)}\\\midrule
Success@1&10&19&14\\
Success@40&27&30&34\\
Ligand-only Top-1&37&34&39\\
Clash at Top-1&5&8&4\\\bottomrule
\end{tabularx}}
\end{table}

Within this diagnostic, union minus baseline is +6.45 points [1.61, 13.95] at Top-1, with four improvements and no regressions; the Top-40 count rises by seven. The selected pose comes from the baseline bank in 45 queries and the guided bank in 17. Full-bank availability at 80 candidates was not tabulated separately. Set inclusion preserves every baseline candidate in the untruncated union, but it does not guarantee preservation in a confidence-ranked prefix or preservation of Top-1 correctness.

\subsection{Selection-only comparator and component diagnostics}
\label{app:posthoc}
The single donor-aware permutation in Equation~\eqref{eq:scoring} was locked before its selected outcomes were computed, after the original historical-guidance results were known. All 200 baseline banks were reranked. Table~\ref{tab:scoring_main} separates the initial, additional, and combined assessable analyses.
\begin{table}[!htbp]
\centering
\caption{Post-hoc selection-only comparison within the stereochemistry checker's input domain. Counts are successes at seed 42 and seed 43. The combined sample contains the initial and additional cohorts. Differences are percentage points with group-bootstrap intervals.}
\label{tab:scoring_main}
\fitbox{\begin{tabularx}{\linewidth}{@{}Yrrr@{}}\toprule
Cohort&No guidance&Donor reranking&Historical donors\\\midrule
Initial (55)&10, 13&15, 18&19, 19\\
Additional (31)&8, 6&10, 8&12, 13\\
Combined (86)&18, 19&25, 26&31, 32\\
\bottomrule\end{tabularx}}
\vspace{0.5ex}
\fitbox{\begin{tabularx}{\linewidth}{@{}Yrr@{}}\toprule
Cohort&Reranking $-$ no guidance&Guidance $-$ reranking\\\midrule

Initial (55)&+9.09 [2.17, 21.25]&+4.55 [$-3.92$, 10.81]\\
Additional (31)&+6.45 [$-1.61$, 15.91]&+11.29 [2.94, 21.43]\\
Combined (86)&+8.14 [2.73, 15.79]&+6.98 [1.28, 12.86]\\
\bottomrule\end{tabularx}}
\end{table}
 The selector reads only the initial ligand graph, supplied historical donors and metal, fixed target entries for those donors, and baseline coordinates/confidence. Native graphs, donor sets, native cardinality, native distances, RMSD, and endpoint labels are excluded from its input process. Whole-input-graph symmetry applies one chirality-preserving permutation to the complete supplied set. Unsupported graphs, empty or invalid contracts, and nonfinite banks would retain the complete saved ranking; none occurred.

The overlap deficit is the first sorting key, followed by descending frozen confidence and ascending sample identifier. Fixed-target radial residuals are recorded as diagnostics but never affect the ordering. The original confidence order uses descending NumPy sorting, and is retained exactly for the no-guidance baseline and any fallback. Every pose remains in the bank, so Success@40 is identical to the baseline on every query and seed. Selection changes on 99 of 200 banks. The input-access record, selected sample IDs, per-pose ranking keys, and paired comparison tables accompany the supplement.

\begin{table}[!htbp]
\centering
\caption{Component outcomes for each method's selected Top-1 pose in the original 62-query cohort. Counts are seed 42 / seed 43, out of 62 at each seed. Radial and directional marginals each allow a complete assignment but need not share a witness.}
\label{tab:components}
\fitbox{\begin{tabularx}{\linewidth}{@{}Yrrrrr@{}}\toprule
Condition&Ligand&Identity&Radial&Direction&Joint\\\midrule
No guidance&37/41&18/20&16/22&27/24&10/13\\
Donor reranking&39/40&37/38&24/28&34/31&15/18\\
Historical donors&34/37&35/35&33/36&23/24&19/19\\
Nearest-$K$&39/41&31/32&30/33&24/26&16/20\\
Native-donor oracle&33/39&39/42&37/42&25/30&21/24\\\bottomrule
\end{tabularx}}
\end{table}

All five conditions preserve supported stereochemistry on 55/62 queries at each seed and sanitize on 62/62. The reranked Top-1 has five fatal-clash cases at each seed; the other conditions' counts for the same 62-query denominator are given in Table~\ref{tab:initial_physical62}. Higher identity accuracy alone is not sufficient for higher joint success: radial/directional geometry and whole-ligand placement must hold on the same selected pose under the joint definition.

We define two baseline-fixed diagnostic subsets separately at each seed. The first contains baseline ligand successes with incorrect donor identity; the second contains baseline ligand successes that fail the joint criterion. Every method is evaluated on those same baseline-defined queries. Table~\ref{tab:baseline_recovery} reports recovered identity with ligand placement retained, joint recovery, and loss of ligand accuracy.

\begin{table}[!htbp]
\centering
\caption{Recovery and regression on baseline-fixed subsets. Entries are seed 42 / seed 43 counts. The denominator differs between the two panels. Identity recovery in the first panel is newly achieved; identity-plus-ligand in the second panel may already hold at baseline and is therefore not reported as a new identity recovery.}
\label{tab:baseline_recovery}
\fitbox{\begin{tabularx}{\linewidth}{@{}Yrrr@{}}\toprule
\multicolumn{4}{@{}l}{Baseline ligand-correct / identity-wrong: 24 / 23 queries}\\\midrule
Condition&\shortstack{Identity repair\\with ligand retained}&Joint recovery&Ligand loss\\\midrule
Donor reranking&11/14&2/6&3/4\\
Historical donors&10/13&6/7&8/4\\
Nearest-$K$&10/9&5/4&3/4\\
Native-donor oracle&13/16&8/8&9/4\\\midrule
\multicolumn{4}{@{}l}{Baseline ligand-correct / joint-wrong: 27 / 28 queries}\\\midrule
Condition&\shortstack{Identity and\\ligand correct}&Joint recovery&Ligand loss\\\midrule
No guidance&3/5&0/0&0/0\\
Donor reranking&14/19&2/6&3/4\\
Historical donors&13/17&7/9&8/5\\
Nearest-$K$&13/14&6/7&3/4\\
Native-donor oracle&16/20&9/10&9/5\\\bottomrule
\end{tabularx}}
\end{table}

\paragraph{Fixed-selector recovery decomposition.}
For the 86-query pooled follow-up, Table~\ref{tab:fixed_selector_availability} summarizes availability and selection under the same donor-aware rule (Section~\ref{sec:fixed_selector}). The 172 paired query--seed observations comprise 21 recoveries, eight regressions, 43 retained successes, and 100 retained failures.
\begin{table}[!htbp]
\centering
\caption{Availability and selection under the same donor-aware rule on 86 assessable queries and two seeds. Counts summarize 172 query--seed observations; each bank contains 40 poses. Differences are historical minus baseline.}
\label{tab:fixed_selector_availability}
\fitbox{\begin{tabularx}{\linewidth}{@{}Yrrr@{}}
\toprule
Outcome & No guidance & Historical & Difference \\
\midrule
Banks with a joint-success pose & 91 & 96 & $+5$ \\
Selected joint successes & 51 & 64 & $+13$ \\
Available but unselected successes & 40 & 32 & $-8$ \\
\bottomrule
\end{tabularx}}
\end{table}
The recoveries admit two descriptive partitions. Ten occur where the baseline bank contains no joint-success candidate, and eleven where one is available. For the sample index selected from the guided bank, the corresponding baseline trajectory fails the joint criterion in 17 recoveries and already succeeds in four. These are alternative partitions of the same 21 recoveries. All 21 corresponding baseline trajectories satisfy the ligand and fatal-clash criteria. Of the 17 baseline failures, three have incorrect realized donor membership; fourteen have correct membership but fail the coordination witness. This distinguishes changes in donor membership from improved realization of an already correct membership. The selected-output comparison also contains eight regressions and the ligand-accuracy loss reported in Section~\ref{sec:fixed_selector}.

\subsection{Group deletion and donor-membership traces}
\label{app:group_trace}
Every frozen group is excluded once in the leave-one-group-out sensitivity analysis (Table~\ref{tab:group_sensitivity}). Removing the group containing the 16 consensus-extension queries leaves 46 initial queries, with historical-minus-baseline and historical-minus-nearest point differences of 14.13 and 2.17 points. These leave-one-group-out estimates are descriptive sensitivity checks.

\begin{table}[!htbp]
\centering
\caption{Ranges of leave-one-group-out paired point estimates in percentage points. They are not confidence intervals. Each row removes every group in the specified cohort once.}
\label{tab:group_sensitivity}
\fitbox{\begin{tabularx}{\linewidth}{@{}Yrrr@{}}\toprule
Cohort&Groups&\shortstack{Historical $-$\\no guidance}&\shortstack{Historical $-$\\Nearest-$K$}\\\midrule
Initial (62)&17&[9.32, 14.13]&[$-0.86$, 2.83]\\
Additional (38)&22&[10.29, 17.14]&[4.41, 8.57]\\
Combined (100)&39&[11.34, 14.29]&[2.08, 4.40]\\\bottomrule
\end{tabularx}}
\end{table}

The saved step-18 and step-19 memberships cover all 8,000 nearest-donor trajectories. All use the historical $K$; the traces confirm that every trajectory retains its chosen set at step 19, as required by the implementation. Table~\ref{tab:membership_trace} compares these actual sets with the historical set using input-only graph symmetry. The donor-element multiset is not constrained and often differs. The controlled comparison consequently changes the atom set and sometimes its element composition, while holding $K$, update timing, operator, and sample count fixed.

\begin{table}[!htbp]
\centering
\caption{Nearest-donor membership from saved trajectories. Counts are seed 42 / seed 43. Set agreement uses whole-input-graph symmetry; element agreement compares multisets. Membership switching is zero throughout.}
\label{tab:membership_trace}
\fitbox{\begin{tabularx}{\linewidth}{@{}Yrrr@{}}\toprule
Cohort&Trajectories/seed&\shortstack{Historical set\\agreement}&\shortstack{Element multiset\\agreement}\\\midrule
Initial&2,480&580/599&1,527/1,563\\
Additional&1,520&395/401&1,055/1,060\\
Combined&4,000&975/1,000&2,582/2,623\\\bottomrule
\end{tabularx}}
\end{table}

\FloatBarrier
\needspace{7\baselineskip}
\subsection{Initial-cohort composition-matched comparison}
\label{app:composition}
This section reports the initial-cohort follow-up separately from the pooled 86-query analysis. The composition-matched control selects the nearest eligible atoms separately within each historical nitrogen, oxygen, and sulfur quota. It retains the selected set through both correction steps and all internal solver iterations. The same parent states, corresponding noise, radial targets, maximum correction iterations, and frozen confidence model are used as for historical guidance. All 124 query--seed banks completed with 40 poses; there were no missing or partial banks. Predictions, traces, technical results, and the evaluator were hash-frozen before endpoint evaluation. Technical checks covered one, two, and three oxygen donors, one sulfur donor, and a one-nitrogen/one-oxygen set.

The initially specified endpoint for this follow-up was historical-minus-composition-matched joint Top-1 success on the 55 inputs assessable by the stereochemistry checker. The 15 represented structural groups are the dependence units. The two seed differences are averaged within each query before resampling groups. Table~\ref{tab:composition_effects} gives the reported paired intervals and query--seed transitions. The original study-wide historical-versus-baseline analysis and the full-cohort results remain separate.

\begin{table}[!htbp]
\centering
\caption{Composition-matched follow-up on 55 queries in 15 groups. Intervals are paired group-bootstrap 95\% intervals; differences are percentage points. Transition counts describe 110 query--seed pairs, not 110 independent queries.}
\label{tab:composition_effects}
\fitbox{\begin{tabularx}{\linewidth}{@{}Yrrr@{}}\toprule
Comparison&Difference&95\% interval&\shortstack{Improvements/\\regressions/ties}\\\midrule
Historical $-$ composition-matched&+2.73&[0.00, 10.17]&7 / 4 / 99\\
Composition-matched $-$ no guidance&+10.91&[4.00, 21.05]&15 / 3 / 92\\
Historical $-$ no guidance&+13.64&[5.26, 26.39]&21 / 6 / 83\\\bottomrule
\end{tabularx}}
\end{table}

The historical-minus-matched difference is +5.45 points at seed 42 and zero at seed 43. Its pooled lower limit is zero, and the prespecified strict-superiority criterion is not satisfied. Including the seven unsupported inputs as automatic failures gives a 62-query sensitivity effect of +2.42 points [0.00, 8.82], with seven improvements, four regressions, and 113 ties. This sensitivity uses the original 17 groups and does not change the conclusion.

\paragraph{Component outcomes.}
Table~\ref{tab:composition_components} separates coordination, whole-ligand placement, and clashes. The coordination-witness indicator, named \texttt{anchor} in the evaluation code, requires joint donor-membership and geometric agreement under one mapping. It is distinct from the identity-only marginal in Table~\ref{tab:components}. Ordinary Full combines ligand success and the coordination witness; joint success further includes stereochemical and clash checks.

\begin{table}[!htbp]
\centering
\caption{Selected Top-1 components and candidate availability in the composition-matched follow-up. Counts pool two seeds over 55 queries, giving 110 query--seed observations. Differences are historical minus composition-matched, in percentage points. These are secondary intervals without multiplicity adjustment.}
\label{tab:composition_components}
\fitbox{\begin{tabularx}{\linewidth}{@{}Yrrrr@{}}\toprule
Endpoint&Historical&\shortstack{Composition-\\matched}&Difference&95\% interval\\\midrule
Coordination witness&42&39&+2.73&[$-2.78$, 9.62]\\
Ligand + coordination witness&38&36&+1.82&[$-3.19$, 9.23]\\
Joint success&38&35&+2.73&[0.00, 10.17]\\
Ligand-only success&69&72&$-2.73$&[$-7.75$, 5.00]\\
Fatal receptor clash&10&15&$-4.55$&[$-8.57$, $-1.14$]\\\midrule
Joint availability at Top-40&59&68&$-8.18$&[$-14.71$, 0.00]\\\bottomrule
\end{tabularx}}
\end{table}

Historical guidance has a lower selected-clash rate than the composition-matched control, but the ligand-only difference is uncertain. Across all 4,400 assessable candidates per condition, clash fractions are 16.25\% for no guidance, 26.73\% for historical guidance, 26.20\% for cardinality-only Nearest-$K$, and 29.57\% for composition-matched Nearest-$K$. Thus both nearest-donor controls and historical correction increase clashes relative to unmodified sampling.

\paragraph{Correctness and dependence.}
Table~\ref{tab:composition_strata} conditions on the native-label audit of historical sets. All seven improvements occur among correct transferred sets, and all four regressions occur among incorrect sets. The retrospectively defined strata share site groups, and both intervals include zero.

\begin{table}[!htbp]
\centering
\caption{Retrospective historical-contract strata. Effects are historical minus composition-matched joint Top-1 success. Groups represented within each stratum are resampled. The strata share some groups.}
\label{tab:composition_strata}
\fitbox{\begin{tabularx}{\linewidth}{@{}Yrrrr@{}}\toprule
Historical set&Queries&Groups&Difference [95\% interval]&\shortstack{Improvements/\\regressions}\\\midrule
Correct&42&11&+8.33 [0.00, 20.83]&7 / 0\\
Incorrect&13&9&$-15.38$ [$-26.19$, 0.00]&0 / 4\\\bottomrule
\end{tabularx}}
\end{table}

\paragraph{Assignment comparison.}
The current 86-query assignment analysis in Section~\ref{sec:assignment_result} uses the trajectory selected by the composition-matched condition's own frozen-confidence ranking. It evaluates that trajectory's assigned set against the native donor annotation, not against the historical condition's independently selected pose. Supplied-set agreement and final coordinate-based donor recovery are different quantities.

\paragraph{Group concentration and availability.}
Only two groups have discordant joint Top-1 outcomes. Group A has three improvements and three regressions; group B has four improvements and one regression. The remaining 13 groups contain only ties. The net three-outcome advantage comes entirely from group B. Group identifiers are retained in the supplementary aggregate table. At Top-40, historical guidance has no improvements and nine regressions relative to the composition-matched control. Greater candidate availability for the nearest-donor policy coexists with slightly higher selected success for historical guidance; neither result alone identifies a causal pathway through confidence or geometry.

\paragraph{Analysis release.}
The matched-control add-on supplies the saved banks, assignment traces,
per-query outcomes, and bootstrap replay for the 86-query assessable domain.
This includes 110 banks from the 55 assessable queries in the initial cohort
and 62 banks from the 31 assessable additional queries. The seven
unassessable queries in each cohort are outside this primary-domain release.
The earlier all-62 aggregates remain a separate historical analysis record. The follow-up's historical-minus-baseline interval [5.26, 26.39] differs slightly from the earlier common-draw reanalysis [5.26, 26.67], despite identical counts. Each interval is kept with its respective analysis.

\subsection{Pooled assignment and baseline-fixed diagnostics}
\label{app:final_recovery}
The pooled follow-up analysis has 86 queries and 33 groups. Composition-matched nearest-donor guidance assigns the native set to 2,688 of 6,880 trajectories. Its own confidence-selected trajectory has an exact assigned set in 107 of 172 query--seed observations. In the diagnostic stratum compatible with the native donor count and elemental quotas, the corresponding denominator is 164 and exactness is 65.24\%. Compatibility is determined with native labels only for analysis; all 172 observations remain in the policy comparison. Across both seeds, 78 queries have at least one exact assignment among their 80 trajectories. This is assignment coverage, not joint-success availability.

The 107 selected exact assignments contain 57 joint successes and 50 failures; the 65 incorrect assignments contain no joint successes. Assigned-set correctness and realized donor membership are not interchangeable: the operator constrains supplied donor distances but contains no non-donor exclusion term.

For the baseline-fixed diagnostic, let $L^a_{n,s}$ and
$I^a_{n,s}$ denote ligand and realized-donor-identity
success on the same selected pose. Joint success is denoted by $R^a_{n,s}$ as defined in Section~\ref{sec:evaluation}. Configuration $0$ denotes the unmodified bank with its original confidence ranking. We form two baseline-defined sets:
\begin{align}
 \mathcal S_{\rm joint}
=
\{(n,s):L^0_{n,s}=1,\ R^0_{n,s}=0\},\\
 \mathcal S_{\rm id}
=
\{(n,s):L^0_{n,s}=1,\ I^0_{n,s}=0\}.
\end{align}
On the assessable domain, $|\mathcal S_{\rm joint}|=69$: historical guidance recovers 26 joint successes and loses ligand accuracy on 14 observations. The narrower $\mathcal S_{\rm id}$ contains 53 observations, with 18 joint recoveries and 12 ligand losses; 29 recover donor identity while retaining ligand accuracy. These quantities are recalculated from the same selected-pose ledger and input-domain mask. Their query--seed counts may include the same query twice, and neither subset is chosen using the intervention outcome.

\paragraph{Available granularity.}
The pooled composition-matched follow-up releases all 86-query seed-by-seed
outcomes, all 172 matched coordinate banks, and both-step donor-assignment
traces. The CPU replay reconstructs coordinate endpoints and the direct
HIST--matched contrasts, including historical-correctness strata and
group-wise paired wins and harms. The two seeds and 40 poses remain repeated
measurements within each query; uncertainty is estimated by resampling the
33 predefined structural groups. The 55-query follow-up tables above therefore remain a separate cohort-specific record rather than a substitute for those pooled inputs. The original replay is supplemented by the coordinate and statistical releases described in Appendix~\ref{app:data_implementation}.

\FloatBarrier
\section{Supporting studies of donor information and pose selection}
\label{app:suppliers}
The following experiments use separate populations and serve as supporting analyses. They are not pooled with the initial or expanded historical-transfer cohorts. The 800-bank replay package covers the historical-transfer experiments, not regeneration of these separate supporting studies.

\subsection{Native-donor control on two docking backbones}
\label{app:oracle}
A separate study evaluates all-40 last-two-step correction with native donor sets on 206 complexes. The second backbone is \textbf{DiffDock-L (v1.1 checkpoints)} \citep{corso2023diffdock,corso2024diffdockl}. Its execution configuration specifies \texttt{model\_\allowbreak dir: ./workdir/v1.1/score\_\allowbreak model} and \texttt{old\_\allowbreak score\_\allowbreak model: false}; recovered generator and confidence checkpoint SHA-256 digests match the official v1.1 distribution. The model release is identified by the recovered configuration and checkpoint digests; contemporaneous container and loaded-weight hashes were not recorded.

Both SurfDock and DiffDock-L improve confidence-selected joint success (Table~\ref{tab:oracle206}). Candidate availability increases by 5.34 points for SurfDock and 10.19 points for DiffDock-L; the latter has a paired 95\% interval of [3.88, 16.50] points. Both backbones were evaluated with native donor sets.
\begin{table}[!htbp]
\caption{Separate 206-complex native-donor experiment with SurfDock and DiffDock-L. DiffDock-L uses the v1.1 generator and confidence checkpoints. Top-1 uses the frozen confidence model; availability is success among all 40 candidates. Intervals are paired 95\% confidence intervals in percentage points.}
\label{tab:oracle206}
\centering
\fitbox{\begin{tabularx}{\linewidth}{@{}Yrrrr@{}}\toprule
Backbone&\shortstack{Baseline\\Top-1 (\%)}&\shortstack{Native donors\\Top-1 (\%)}&\shortstack{Difference\\(points)}&95\% interval\\\midrule
SurfDock \citep{cao2025surfdock} \venue{Nat. Methods '25}&27.67&49.03&$+21.36$&$[13.59, 29.13]$\\
DiffDock-L \citep{corso2024diffdockl} \venue{ICLR '24}&14.56&20.39&$+5.83$&$[0.97, 10.68]$\\\midrule
\multicolumn{5}{@{}l}{Availability: SurfDock 67.96\% $\rightarrow$ 73.30\%; DiffDock-L 34.47\% $\rightarrow$ 44.66\%.}\\\bottomrule
\end{tabularx}}
\end{table}

\subsection{Learned donor sets and downstream effects}
An earlier experiment uses a structured ligand-graph and metal-element donor predictor with approximately 4.87 million parameters. On 104 queries and two seeds, its output is passed to the same class of last-two-step distance correction. The mean selected joint-success rates are 25.96\% without guidance, 25.00\% with predicted donors, and 41.35\% with native donors. The native-versus-baseline difference is +15.38 points [3.75, 24.04].

Retrospective exactness separates opposing effects. On 102 query--seed banks with correct predicted sets, baseline success is 33.33\% and predicted-donor success is 46.08\%, a +12.75-point change. On 106 banks with incorrect sets, success changes from 18.87\% to 4.72\%, a $-14.15$-point change. Thus the donor predictor's average accuracy does not by itself describe the balance of improvement and degradation under intervention. These 208 banks are repeated evaluations of 104 queries, not 208 distinct biological samples.

\subsection{External pretraining of a donor predictor}
A separate donor-prediction comparison uses the same structured architecture and target evaluation while varying ligand-encoder pretraining. The target-only baseline reaches 35.93\% complete-set exactness, a matched externally pretrained model 37.27\%, and a variant using a listwise complete-set pretraining objective 36.37\%. The listwise variant differs from the matched pretrained model by $-0.90$ points, with a component-bootstrap interval of [$-2.04$, 0.34]. Same-cardinality wrong-set errors rise from 29.78\% to 31.65\%.

The external corpus contains 70,069 records, of which 69,997 pass label and atom-mapping checks. Compatibility filtering gives 41,227 records, and exact-ligand/scaffold exclusions against the target collections leave 33,837. The listwise pretraining variant did not improve complete-set prediction over the matched pretrained baseline.

\subsection{Historical-only evidence screening}
\label{app:historical_screen}
We also evaluated a stricter eligibility rule using only earlier reference records. Each historical record is treated as a pseudo-query, with transfer restricted to still earlier structures; canonical ligand graphs are deduplicated. A family qualifies only if it has at least three distinct historical graph units and no observed transfer errors. There are 1,146 pseudo-query trials, of which 369 are assessable, and no current query qualifies.

Among the 33 queries covered by strict site-matched transfer, twelve have no assessable historical graph and twelve have only one. No zero-error family has two or more independent evidence units. Families with four and nine evidence units have historical precision of 75\% and 22.22\%, respectively. The historical-only screen is distinct from the broader transfer rule used in the docking experiments. The rule's zero coverage reflects limited consistent historical evidence.

\subsection{Chemical-descriptor models for complete donor sets}
\label{app:classical}
Three classical learning methods test whether chemical and pocket descriptors improve the frozen MetalloDock donor-head output. The shared development data contain 339 queries in 172 biological clusters and 727,650 candidate sets of eligible nitrogen, oxygen, and sulfur atoms, with cardinality one to four. Five outer folds and fixed inner splits are reused. Labels use whole-graph chirality-preserving donor-set equivalence, and only out-of-fold predictions enter the result tables.

The primary metric is cluster-macro exactness: within-cluster query accuracy is averaged equally over 172 clusters. The adjacent case counts instead weight all 339 queries equally. Paired percentile intervals resample the 172 cluster means 10,000 times with NumPy RandomState seed 20260922. These settings belong to this development study, not to the historical replay.

Shared descriptors describe atom chemistry, selected donor composition, pair topology, metal-centered protein environments, pocket--set interactions, and Gasteiger partial-charge summaries \citep{gasteiger1980}. Retained water information is absent in all 339 inputs. Gasteiger calculation fails on 14 queries, represented with missing-value flags rather than zero charge. The receptors are ligand-removed holo structures; pretrained-model exposure and protein-family independence are unresolved.

\paragraph{Models.}
The CatBoost \citep{prokhorenkova2018catboost} hybrid uses a depth-four
PairLogit ranker with 333 columns, including MetalloDock scores and
outputs of the learned complete-set predictor. Training uses at most 500 iterations, learning rate 0.03, L2 coefficient 10, and normalized discounted cumulative gain at rank two for early stopping with patience 50. The log-linear model adds a 153-feature structured score to a fixed, train-standardized MetalloDock independence offset. It uses cluster-weighted catalogue cross-entropy with L2 coefficients in $\{0.1,1,10\}$. The kernel model adds a Ridge residual with a 128-component Nystr\"om radial-basis approximation. Its charge-augmented version uses 322 features; the charge-free version uses 316, with training-only imputation and scaling. Inner cluster-macro Top-1 chooses regularization for both models.

\begin{table}[!htbp]
\caption{Prespecified chemical-descriptor comparisons against the MetalloDock expected-count decoder. Differences and 95\% intervals refer to cluster-macro exactness.}
\label{tab:ml_primary}
\centering
\fitbox{\begin{tabularx}{\linewidth}{@{}Yrrrr@{}}\toprule
Method&\shortstack{Cluster\\exact (\%)}&\shortstack{Exact queries\\($n/N$)}&\shortstack{Difference\\(points)}&95\% interval\\\midrule
MetalloDock expected-count \citep{zhang2026metallodock} \venue{JACS '26}&45.49&150/339&---&---\\
CatBoost hybrid \citep{prokhorenkova2018catboost} \venue{NeurIPS '18}&45.31&140/339&$-0.18$&[$-6.87$, 6.56]\\
Log-linear full model&46.21&125/339&+0.72&[$-5.87$, 7.13]\\
Charge-augmented kernel&41.44&114/339&$-4.05$&[$-8.50$, 0.30]\\\bottomrule
\end{tabularx}}
\end{table}

All three primary intervals include zero (Table~\ref{tab:ml_primary}). In the log-linear model, adding donor-pair features improves its unary/cardinality variant by +7.25 points [1.54, 13.13], while adding pocket interaction gives +0.57 [$-4.14$, 5.34]. The pair block also changes the cardinality representation through redundant count-derived features. Adding the Gasteiger block to the kernel changes exactness by $-0.58$ [$-3.49$, 2.33] points; it also changes the kernel distance and bandwidth. Removing added pocket descriptors leaves any protein information already encoded in the frozen predictor.

Compared with MetalloDock expected-count, CatBoost gains 37 correct queries and loses 47; log-linear gains 35 and loses 60; the kernel gains five and loses 41. These earlier chemical-descriptor runs use RDKit 2023.03.1 \citep{rdkit2023}; the current saved-bank analysis uses release 2025.09.6. A post-hoc restriction to the MetalloDock-predicted donor count raises the log-linear pair/full variants to 47.26/47.73\% cluster-macro exactness, with 150/339 and 152/339 correct queries. The respective baseline differences are +1.77 [$-2.83$, 6.29] and +2.24 [$-1.50$, 6.13] points. These exploratory decodings were not used for the historical docking experiments. The model comparison jointly varies the learner, prior usage, feature representation, and training loss.

\subsection{Selection from mixed candidate banks}
An exploratory study trains a metal-local ranker on a separate 192-complex development collection and evaluates it on 104 queries. It combines 40 unmodified and 40 predicted-donor-guided poses, retains the common-confidence top eight, and chooses one with a learned metal-local ranker. The feature set includes confidence, rank, contact/clash and stereochemistry features, plus metal-distance, cardinality, and overcoordination summaries. Common confidence on the mixed bank gives 27.40\% joint success, the ranker on an uncorrected bank 35.58\%, and the ranker on the mixed bank 36.06\%. The mixed-versus-common-confidence difference is +8.65 points [1.49, 13.54], but the difference from the uncorrected-bank ranker is only +0.48 points. This supports the usefulness of metal-local selection features without isolating a benefit from adding corrected candidates. The main historical-transfer experiment keeps the original confidence model fixed.

\FloatBarrier
\section{Earlier controller configuration}
\label{app:earlier}
Before the last-two-step radial correction, we evaluated a different controller that branches from a step-five SurfDock state and applies corrections after steps 6, 8, 10, 12, 14, 18, and 19. This controller uses one damped Gauss--Newton update per call. Its residual combines donor distance targets, movement from the incoming pose, and non-donor ordering. With $d_i=\|\mathbf x_i-\mathbf z_m\|_2$, its ordering term is proportional to
\begin{equation}
 \max\left\{0,\ \max_{i\in D}d_i+0.20\ang-\min_{j\notin D}d_j\right\}.
\end{equation}
Non-members include all ligand heavy atoms outside the
supplied set. Let $r_{\rm old}$ denote this controller's
combined distance, movement, and ordering residual, and
$J_{\rm old}$ its Jacobian. The linear step solves
\begin{equation}
 (J_{\rm old}^{\top}J_{\rm old}+0.10I)\Delta q
=
-J_{\rm old}^{\top}r_{\rm old}.
\end{equation}
It uses separate translation and angular caps in addition to a Cartesian RMS cap. A trial that increases the count of protein contacts below the specified threshold is rejected. This differs from the radial-only last-two-step operator, which has no contact guard, movement residual, or exclusion residual.

Earlier branch evaluations use a shared bank of 16 unmodified poses together with 16 additional poses from a correction condition. Their primary endpoint measures whether at least one successful pose occurs in the union. The main experiment instead applies a last-two-step operator to all 40 trajectories and evaluates the confidence-selected Top-1 pose. Preserving an original bank changes the possibility of losing baseline support, while selecting the best-ranked pose asks a different question from availability. These designs are therefore kept separate when interpreting the method.

In a 50-case scheduling comparison, the unmodified condition has eleven successes, fixed membership and an open-loop variant each have 23, and two adaptive variants have 20 and 22. Fixed membership matches the highest observed success count in this 50-case comparison. Several solver, timing, and penalty settings differ across the controller configurations.

\FloatBarrier
\section{Contract accuracy and intervention benefit}
\label{app:utility}
For a fixed evaluation population, correction operator,
selector, and endpoint, let $C=1$ indicate that a
historical donor contract is supplied. On these queries,
let $Z=\mathbf 1\{\widehat D\equiv D^\star\}$ denote
agreement with the native donor set.

Write $\delta_n$ for query $n$'s two-seed paired change
as in Equation~\eqref{eq:paired}, with the contrast
suppressed, and $\delta$ for the corresponding random-query quantity. Define
\[
\begin{aligned}
p&=\Pr(Z=1\mid C=1),\\
g&=\mathbb E[\delta\mid Z=1,C=1],\\
\ell&=\mathbb E[\delta\mid Z=0,C=1].
\end{aligned}
\]
Here $p$ is the exact-contract probability among supplied
queries, rather than supplier coverage. Both $g$ and $\ell$ are signed mean paired changes in success probability, rather than absolute success rates.

The law of total expectation gives
\begin{equation}
 \Delta
:=
\mathbb E[\delta\mid C=1]
=
pg+(1-p)\ell.
 \label{eq:utility}
\end{equation}
Thus, the overall mean paired change is the weighted
average of the changes in the correct- and incorrect-contract strata. The weights are their query proportions and must match the overall estimand.

For compact notation, let $\widehat\Delta^{(N)}_{H-b}$ denote the empirical contrast of historical guidance $H$ against condition $b$ on $N$ queries. Here $0$ is no guidance, $\mathrm{near}$ is the count-only nearest policy, and $\mathrm{comp}$ is the composition-matched policy. The following examples use frozen-confidence selection.

For the main assessable population,
\begin{equation}
 \widehat\Delta^{(86)}_{H-0}
=
\frac{69}{86}\cdot\frac{30}{2\cdot69}
+
\frac{17}{86}\cdot\frac{-4}{2\cdot17}
=
\frac{26}{172}
\approx0.1512.
\end{equation}
The weights give the proportions of correct and incorrect
historical contracts. Within each stratum, the numerator
is the net difference in successful query--seed outcomes,
and the denominator is twice the query count because
each query has two seeds.

\begin{figure}[!htbp]
\centering
\includegraphics[width=\linewidth]{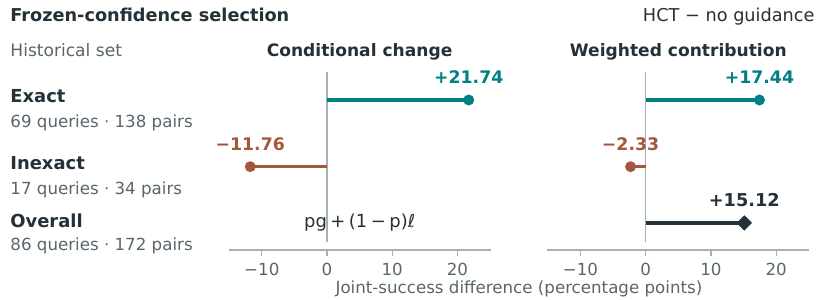}
\caption{Conditional paired effects and query-weighted intervention benefit in SurfDock.
The left plot shows conditional changes in joint success for HCT guidance against no guidance; the right plot shows their query-weighted contributions and the overall contrast $\Delta=pg+(1-p)\ell$. Both plots use the same percentage-point scale and frozen-confidence selection.
Historical donor membership is classified retrospectively using native donor labels. Exact membership comprises 69 queries (138 query--seed pairs), with a net gain of 30 joint successes ($g=+21.74$ percentage points); inexact membership comprises 17 queries (34 pairs), with a net loss of 4 ($\ell=-11.76$ percentage points).
The weights $p=69/86$ and $1-p=17/86$ give contributions of $+17.44$ and $-2.33$ percentage points, summing to $\Delta=+15.12$ percentage points ($26/172$).
Native labels define evaluation strata only. These descriptive means have different stratum denominators; no confidence intervals are drawn. The fixed donor-aware comparison appears in Figure~\ref{fig:bank_selection}(b).}
\label{fig:contract_accuracy}
\end{figure}

The retained earlier analyses give
\begin{align}
 \widehat\Delta^{(62)}_{H-0}
=
\frac{45}{62}\cdot\frac{18}{90}
+
\frac{17}{62}\cdot\frac{-3}{34}
=
\frac{15}{124},\\
 \widehat\Delta^{(62)}_{H-\mathrm{near}}
=
\frac{45}{62}\cdot\frac{6}{90}
+
\frac{17}{62}\cdot\frac{-4}{34}
=
\frac{2}{124},\\
 \widehat\Delta^{(100)}_{H-\mathrm{near}}
=
\frac{73}{100}\cdot\frac{11}{146}
+
\frac{27}{100}\cdot\frac{-4}{54}
=
\frac{7}{200},\\
 \widehat\Delta^{(55)}_{H-\mathrm{comp}}
=
\frac{42}{55}\cdot\frac{7}{84}
+
\frac{13}{55}\cdot\frac{-4}{26}
=
\frac{3}{110}.
\end{align}
These identities decompose the observed paired effects by contract correctness. The 55-query composition contrast is distinct from the pooled follow-up 86-query comparison.

Let $c\ge0$ be the expected incremental intervention cost, expressed on the same utility scale as the success-probability change. Then $U=pg+(1-p)\ell-c$. Both $g$ and $\ell$ may have either sign. For the threshold calculation, consider $\ell\le0$ and write $h=-\ell\ge0$. If $g+h>0$, then \[ U>0 \quad\Longleftrightarrow\quad p>\frac{h+c}{g+h}. \]
This threshold depends on the supplier's actual error
distribution. Per-pose clash fractions, computation time,
and per-query success changes must be expressed on a common utility scale before they can be combined.

Subscripts $L$ and $U$ denote lower and upper bounds,
respectively. Suppose that the following bounds hold
simultaneously: $p\ge p_L$, $g\ge g_L$, $h\le h_U$, and $c\le c_U$, with $g_L+h_U\ge0$. Then
\begin{equation}
 U\ge p_Lg_L-(1-p_L)h_U-c_U.
\end{equation}
Indeed, substitution bounds $U$ by $p(g_L+h_U)-h_U-c_U$, which is nondecreasing in $p$. Separate marginal 95\% intervals do not imply the required simultaneous event. Using the bound as a selection rule requires separate calibration and validation \citep{bates2021risk,angelopoulos2025ltt}.

\begin{sloppypar}For the original roster only, unchanged baseline outputs outside the 62 supplied queries would imply $(62/304)(15/124)=\allowbreak 15/608\approx0.0247$. This accounting identity does not estimate the pooled 86-query policy's population coverage or provide a full-roster confidence interval.\end{sloppypar}

\clearpage
\section{Molecular examples}
\label{app:molecular}
\begin{figure}[H]
\centering
\includegraphics[width=\linewidth]{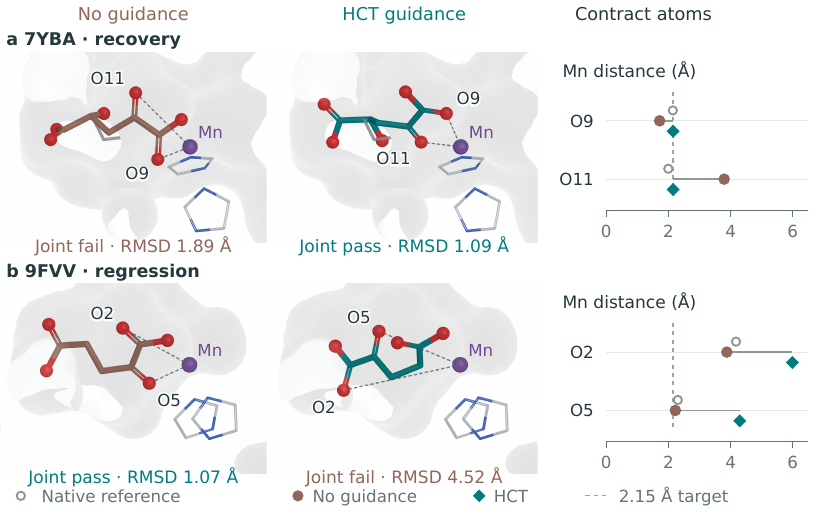}
\caption{\textbf{Contract membership and observed metal--donor geometry.}
Each selected output is overlaid on the native ligand (thin gray), used only for evaluation.
\textbf{(a)} Q023, 7YBA IO9 A802/Mn A801: no-guidance/HCT samples 36/32; native and historical donors O9/O11.
\textbf{(b)} Q002, 9FVV AKG A801/Mn A805: samples 19/10; native donors O3/O5, historical donors O2/O5.
All outputs are independently frozen-confidence-selected at seed 43, from distinct trajectories; the original size/ID case-selection rule above is retained.
Atom labels are one-based input-graph indices, mapped to zero-based arrays in the manifest.
Right-hand plots measure Mn distances to the historical-contract atoms in native, no-guidance and HCT coordinates on one 0--6.5\,\AA\ scale.
The native O2 point in (b) evaluates the historical atom's reference position; O2 is not a native donor.
The dashed 2.15\,\AA\ line is the fixed Mn--O target; structural connectors identify measured distances rather than inferred bonds.
The recovery reaches approximately 2.15\,\AA, while the regression retains 5.99/4.29\,\AA: assigned membership need not be realized.
Labels reproduce stored ligand RMSD and joint-success flags. Recorded fatal-clash flags are zero for all four outputs.
Native geometry uses open gray circles, no guidance brown circles, and HCT teal diamonds.
Receptor context and camera/crop are fixed within each case; coordinates are unchanged, without alignment or minimization.
These examples use frozen-confidence selection; the fixed donor-aware comparison is shown in Figure~\ref{fig:bank_selection}(b).
Exact distances, inputs, sample IDs, atom mapping and cameras are archived with the editable figure.}
\label{fig:molecular}
\end{figure}

Figure~\ref{fig:molecular} illustrates a successful change and a regression in the initial cohort. Examples were selected retrospectively, taking the smallest ligand by heavy-atom count and then query identifier within each of two seed-43 categories: a joint-success improvement with a correct historical set, and a regression with an incorrect set of the correct cardinality. The rule selects small ligands for visual clarity. The plotted predictions are each method's selected Top-1 pose; they need not be the same trajectory index.

The supplementary renderer reads the stored coordinate arrays and native graphs directly. Provenance for Figure~\ref{fig:molecular}, including query identifiers, selected samples, donor indices, camera parameters, and input hashes, is archived with the editable figure sources. The plotted coordinates are exported without modification.

\subsection{Structural provenance for Figure~\ref{fig:overview}}
\label{app:figure1_provenance}

Figure~\ref{fig:overview} uses query Q023, PDB 7YBA, ligand IO9 A802 and Mn A801. HCT maps donor membership from references 6YYV/6YYX (ligand Q1Z). The ribbon rendering shows the ligand-binding region of the 6YYV chain-A complex. The query graph uses one-based atom labels: HCT supplies O9/O11 (zero-based 8/10), and the trajectory-local nearest example supplies O7/O10 (zero-based 6/9). The latter assignment is recorded at seed 43, sample 0, immediately after native update 18 and is retained through update 19. All eligible atoms in this example are oxygen, so the count-only and composition-matched nearest constraints coincide.

The structural pair shows independently frozen-confidence-selected SurfDock outputs at seed 43: no-guidance sample 36 and HCT-guided sample 32. Their Mn--O9/O11 distances are 1.72/3.79\,\AA\ and 2.15/2.15\,\AA, respectively. Dashed connectors mark these measured distances. The operator uses a fixed Mn--O radial target of 2.15\,\AA\ and residual normalization $\sigma=0.10$\,\AA. Both poses use the same camera, crop, and receptor cutaway to expose the binding site. Protein, metal, and ligand coordinates are unchanged. Their ligand RMSDs and joint-success outcomes are shown in Figure~\ref{fig:bank_selection}(a), together with the regression example detailed in Appendix~\ref{app:molecular}.

\FloatBarrier
\section{Follow-up comparisons and execution details}
\label{app:followups}
The follow-up studies reuse the 86-query assessable domain and its 33 structural groups. They address selection, correction timing, physical quality, and backbone transfer rather than introducing a new biological validation cohort. Paired intervals in this section use the common analysis in Appendix~\ref{app:statistics}; historical cohort-specific intervals are retained in Appendix~\ref{app:fullcohort}. MetalloDock contributes one output per query and is reported separately.

\subsection{Information policies and fixed-selector effects}
\label{app:followup_policy}
\begin{table}[!htbp]
\centering
\caption{Joint Top-1 success in the pooled follow-up (86 queries, 33 groups, two seeds). Counts summarize 172 query--seed outcomes from 40-pose banks. Donor-aware reranking is a post-hoc selection-only comparison.}
\label{tab:main86_counts}
\fitbox{\begin{tabularx}{\linewidth}{@{}Yrr@{}}\toprule
Condition&Successful query--seed outcomes&Mean success (\%)\\\midrule
No guidance&37/172&21.51\\
Nearest-$K$ (count only)&56/172&32.56\\
Composition-matched Nearest-$K$&57/172&33.14\\
Historical donors&63/172&36.63\\
Native-donor oracle&70/172&40.70\\\midrule
Donor-aware reranking&51/172&29.65\\\bottomrule
\end{tabularx}}
\end{table}

\begin{table}[!htbp]
\centering
\caption{Paired joint Top-1 differences in percentage points. Effects average two seeds within each query; 95\% intervals resample structural/site groups. Reranking comparisons are post-hoc. Bootstrap settings are given in Appendix~\ref{app:statistics}.}
\label{tab:main86_effects}
\fitbox{\begin{tabularx}{\linewidth}{@{}Yrr@{}}\toprule
Comparison&Difference&95\% confidence interval\\\midrule
Historical donors $-$ no guidance&$+15.12$&$[8.00,24.63]$\\
Nearest-$K$ (count only) $-$ no guidance&$+11.05$&$[5.49,17.86]$\\
Composition-matched Nearest-$K$ $-$ no guidance&$+11.63$&$[6.12,18.84]$\\
Native-donor oracle $-$ no guidance&$+19.19$&$[11.21,28.51]$\\
Historical $-$ Nearest-$K$ (count only)&$+4.07$&$[0.00,10.37]$\\
Historical $-$ composition-matched&$+3.49$&$[-0.42,8.93]$\\
Native-donor oracle $-$ composition-matched&$+7.56$&$[1.52,14.29]$\\\midrule
Donor-aware reranking $-$ no guidance&$+8.14$&$[2.73,15.79]$\\
(Historical + confidence) $-$ (Baseline + donor-aware reranking)&$+6.98$&$[1.28,12.86]$\\\bottomrule
\end{tabularx}}
\end{table}

Historical and trajectory-local contracts are different inputs to the same reference correction. The nearest-donor policies use historical cardinality, and composition matching also uses its elemental quotas. The native-donor condition can correct both these quantities as well as atom membership.

\subsubsection{Selector definition and bank intervention}
\label{app:followup_selector}
The fixed donor-aware selector ranks each bank independently by Equation~\eqref{eq:scoring}, then by descending confidence and ascending sample ID. Both 6,880-row permutations on the assessable subset reproduce the stored rankings when access to native graphs, coordinates, and endpoint labels is blocked. Its C0 and historical-guided joint counts are 51/172 and 64/172, with 21 improvements, eight regressions, 43 retained successes, and 100 retained failures.

A separate selector prioritizes finite candidates by: (i) absence of receptor contacts below 1.2~\AA; (ii) whether every donor's absolute target-distance residual is at most 0.5~\AA; (iii) the maximum absolute residual; (iv) decreasing confidence; and (v) sample ID. For its count and composition variants, donors minimize absolute target-distance residual rather than the raw distance used in the sampling-time nearest controls. The historical variant uses the transferred indices. The rule falls back to the original confidence choice when every candidate is nonfinite. Native-input access is blocked during replay.
\begin{table}[!htbp]
\centering
\caption{Radial/clash selectors on 86 queries. Each row applies its selector independently to the common unmodified baseline bank and the bank generated by the listed guidance policy. Entries count joint successes among 172 query--seed outcomes. Differences are guided minus baseline, with nominal 95\% group-bootstrap intervals in percentage points.}
\label{tab:radial_selectors}
\fitbox{\begin{tabularx}{\linewidth}{@{}YYrrr@{}}
\toprule
Selector donor information & Guidance policy
& Baseline & Guided & Guided $-$ baseline \\
\midrule
Count
& Count-only Nearest-$K$
& 32 & 16 & $-9.30 [-18.69, -1.41]$ \\

Count and composition
& Composition-matched Nearest-$K$
& 39 & 29 & $-5.81 [-15.38, 3.47]$ \\

Historical membership
& Historical-guided CICG
& 48 & 38 & $-5.81 [-15.13, 3.80]$ \\
\bottomrule
\end{tabularx}}
\end{table}

For the historical-membership condition, the radial/clash selector yields 48/172 joint successes on the unmodified bank and 38/172 on the historical-guided bank. The overlap-based selector yields 51/172 and 64/172 on these same banks, respectively.

The fixed-selector initial 55 queries give 33/110 versus 40/110 successes (+6.36 points [$-3.00$, 14.29]); the additional 31 give 18/62 versus 24/62 (+9.68 [0.00, 20.31]). The full 100-query sensitivity retains the same success counts, 51/200 versus 64/200 (+6.50 [0.90, 12.07]), counting unsupported inputs as failures. On 86 queries, ten original groups contribute positive net changes, three negative changes, and twenty zero changes. The retained leave-one-group-out point effects range from +6.10 to +9.26 points. The reference-linked sensitivity merges the 33 groups into 23, giving +7.56 points [1.14, 14.62]. This sensitivity concerns the fixed-selector contrast and does not establish protein-family independence.

The 21 recoveries comprise ten observations lacking any baseline joint-success candidate and eleven with available baseline support. For the guided-selected sample ID, seventeen corresponding baseline trajectories fail and four already succeed. Of those seventeen witness recoveries, three also recover donor identity; fourteen already have correct identity but fail radial or directional agreement. Among the eight regressions, five selected guided outputs fail ligand placement and six fail the coordination witness, with three failing both. None of those eight outputs has a fatal clash; other observations still contribute to the aggregate clash increase.

\subsection{Terminal-only timing comparison}
\label{app:followup_timing}
The timing hypothesis requires interleaving to improve joint Top-1 by at least three percentage points over terminal placement, with a positive group-bootstrap lower limit. In the reference SurfDock run, one correction follows native update 18 and a second follows update 19. The terminal condition defers both calls until after update 19. They are two separate calls, each with at most five relinearizations, rather than one ten-iteration solve. The second call rebuilds its reference conformer and zero intrinsic coordinates from the first call's output. The numerical operator, fixed targets, caps, and maximum call count are otherwise unchanged, and new terminal poses are rescored.

For the nearest-donor timing control, the same-donor variant selects at the saved post-update-18 pose, defers correction, and reuses that set in both terminal calls. A separate terminal-reassigned variant selects again at the post-update-19 pose; it changes 362 of 6,880 memberships and therefore alters assignment as well as timing. All 516 terminal banks retain their 40 candidate slots. The audit checks 20,640 saved prefix and donor records and 1,548 pre-run source hashes. Post-prefix RNG/noise tensors and the terminal scorer's historical loaded-weight hash were not recorded.
\begin{table}[!htbp]
\centering
\caption{Timing comparison on 86 queries. Differences are interleaved minus terminal-only, with common group-bootstrap 95\% intervals. Counts use 172 query--seed observations; candidate percentages use 6,880 poses per condition. Both terminal calls have at most five relinearizations.}
\label{tab:timing_followup}
\fitbox{\begin{tabularx}{\linewidth}{@{}Yrrr@{}}\toprule
Policy or outcome & Interleaved & Terminal-only & Difference [95\% interval] \\\midrule
Historical: joint success & 63 & 65 & $-1.16 [-3.19, 1.15]$ \\
Historical: ligand success & 101 & 105 & $-2.33 [-6.25, 1.72]$ \\
Historical: selected clash & 22 & 18 & $+2.33 [0.50, 4.84]$ \\
Historical: candidate clash (\%) & 27.22 & 30.99 & $-3.76 [-4.94, -2.45]$ \\
Count-only, fixed set: joint & 56 & 59 & $-1.74 [-3.52, 0.00]$ \\
Count-only, reassigned set: joint & 56 & 59 & $-1.74 [-3.52, 0.00]$ \\
\bottomrule\end{tabularx}}
\end{table}

The direct historical interval is [$-3.19$, 1.15] points, whose upper endpoint is below the three-point superiority target. Terminal-only has a slightly higher observed joint count and fewer selected clashes, but more clashes across its entire bank. The reference schedule remains unchanged. Counts and limits define the matched maximum budget; accepted iterations, backtracking, and wall times are retained separately in \texttt{TERMINAL\_\allowbreak OPERATOR\_\allowbreak COSTS.tsv}.

\subsection{Physical checklist and joint efficacy}
\label{app:followup_pb}
The stored PoseBusters evaluation covers 27,520 poses in 688 banks: no guidance, historical donors, count-only nearest donors, and composition-matched nearest donors. 
Define $Q(y)\in\{0,1\}$ as the conjunction, for pose $y$,
of eight internal-quality checks:
\texttt{sanitization},
\texttt{all\_\allowbreak atoms\_\allowbreak connected},
\texttt{bond\_\allowbreak lengths},
\texttt{bond\_\allowbreak angles},
\texttt{internal\_\allowbreak steric\_\allowbreak clash},
\texttt{aromatic\_\allowbreak ring\_\allowbreak flatness},
\texttt{double\_\allowbreak bond\_\allowbreak flatness}, and
\texttt{internal\_\allowbreak energy},
together with
\texttt{volume\_\allowbreak overlap\_\allowbreak with\_\allowbreak protein}.
Missing flags are treated as false. Native identity, chirality, and RMSD columns from the redock invocation are retained as outputs but excluded from $Q(y)$. The joint endpoint remains $R(y)$ as defined in Section~\ref{sec:evaluation}; their intersection on the same pose is $R(y)Q(y)$.

The 86 receptor files contain no metal, water, or other cofactor atom records. The target metal is separate guidance metadata, so $Q(y)$ assesses protein overlap rather than metal or solvent contacts. Completed outputs contain no remaining error rows after technical retries. The selected records agree with the corresponding rank-one rows of the full-bank files for all nine checklist columns and sample IDs. The exact historical PoseBusters version and default-threshold YAML were not pinned; the release reproduces stored-check aggregation rather than asserting an exact re-execution of that checker environment.
\begin{table}[!htbp]
\centering
\caption{Confidence-selected joint and stored-checklist outcomes on 86 queries (172 observations per condition). The checklist is the nine-flag conjunction specified in the text.}
\label{tab:pb_selected_complete}
\fitbox{\begin{tabularx}{\linewidth}{@{}Yrrr@{}}\toprule
Condition & Joint & Checklist & Joint and checklist \\\midrule
No guidance & 37 & 140 & 33 \\
Historical donors & 63 & 117 & 54 \\
Count-only Nearest-$K$ & 56 & 121 & 49 \\
Composition-matched Nearest-$K$ & 57 & 116 & 49 \\
\bottomrule\end{tabularx}}
\end{table}

\begin{table}[!htbp]
\centering
\caption{Stored PoseBusters pass fractions across all 6,880 candidates per condition (\%). Each query--seed bank contributes 40 poses. This aggregation is separate from the confidence-selected comparison.}
\label{tab:pb_allposes_complete}
\fitbox{\begin{tabularx}{\linewidth}{@{}Yrrr@{}}\toprule
Condition & Internal & Protein overlap & Checklist \\\midrule
No guidance & 80.84 & 71.66 & 59.49 \\
Historical donors & 80.87 & 45.84 & 37.76 \\
Count-only Nearest-$K$ & 80.73 & 51.12 & 42.67 \\
Composition-matched Nearest-$K$ & 80.93 & 47.83 & 39.87 \\
\bottomrule\end{tabularx}}
\end{table}

For each bank we compute its 40-pose checklist fraction, average seeds within query, and resample the original groups. Historical guidance changes the all-candidate checklist fraction by $-21.73$ points [$-26.87$, $-15.31$]. Its eight-check internal-quality conjunction changes by +0.03 [$-0.38$, 0.55], whereas protein-overlap compliance changes by $-25.81$ [$-31.69$, $-18.77$]. Thus the stored all-pose quality loss is concentrated in protein overlap, not in a demonstrated systematic loss of internal geometry. These additional flags remain secondary outcomes.

\begin{table}[!htbp]
\centering
\caption{Partition of confidence-selected outputs on 86 queries. This is a within-condition partition, not a paired transition table.}
\label{tab:checklist_partition}
\fitbox{\begin{tabularx}{\linewidth}{@{}Yrr@{}}\toprule
Selected-output state & No guidance & Historical donors \\\midrule
Joint success; checklist pass & 33 & 54 \\
Joint success; checklist fail & 4 & 9 \\
Joint failure; checklist pass & 107 & 63 \\
Joint failure; checklist fail & 28 & 46 \\\midrule
Total & 172 & 172 \\\bottomrule
\end{tabularx}}
\end{table}
\begin{table}[!htbp]
\centering
\caption{Geometric outcomes for five sampling conditions on 86 queries.
Selected counts summarize 172 query--seed outcomes;
candidate fractions use 6,880 poses per condition. }
\label{tab:main86_physical}
\fitbox{\begin{tabularx}{\linewidth}{@{}Yrrrr@{}}\toprule
Condition&\shortstack{Ligand\\Top-1}&\shortstack{Top-1\\clashes}&\shortstack{All-pose\\clash (\%)}&\shortstack{All-pose geometry-\\valid (\%)}\\\midrule
No guidance&106&11&16.83&83.17\\
Historical donors&101&22&27.22&72.78\\
Nearest-$K$ (count only)&109&20&26.32&73.68\\
Native-donor oracle&101&21&27.17&72.83\\
Composition-matched Nearest-K & 105 & 22 & 29.39 & 70.61 \\\bottomrule
\end{tabularx}}
\end{table}

\begin{table}[!htbp]
\centering
\caption{Confidence-ranked joint success on 86 queries. Counts use 172 observations; all banks contain 40 poses. The complete bank is independent of which post-hoc selector subsequently permutes it.}
\label{tab:topk86_followup}
\fitbox{\begin{tabularx}{\linewidth}{@{}Yrrrrr@{}}\toprule
Condition & Top-1 & Top-2 & Top-4 & Top-8 & Top-40 \\\midrule
No guidance & 37 & 50 & 70 & 83 & 91 \\
Count-only Nearest-$K$ & 56 & 73 & 88 & 101 & 106 \\
Composition-matched Nearest-$K$ & 57 & 76 & 91 & 103 & 106 \\
Historical donors & 63 & 75 & 84 & 93 & 96 \\
Native-donor oracle & 70 & 85 & 93 & 103 & 106 \\
\bottomrule\end{tabularx}}
\end{table}

Historical-minus-composition-matched Top-40 availability is $-5.81$ points [$-10.82$, 0.00], with one improvement and eleven regressions across query--seed observations. Higher selected historical success coexists with lower bank support. Donor-aware selection preserves each bank's availability because it only permutes candidates.

\subsection{Historical guidance in the DiffDock-L follow-up}
\label{app:followup_diffdock}
The technical rerun uses the DiffDock-L v1.1 score-model configuration (\texttt{old\_\allowbreak score\_\allowbreak model=false}) and its configured confidence model \citep{corso2024diffdockl}. The wrapper specifies \texttt{best\_\allowbreak ema\_\allowbreak inference\_\allowbreak epoch\_\allowbreak model.pt} for generation and \texttt{best\_\allowbreak model\_\allowbreak epoch75.pt} for confidence. Recovered checkpoint hashes and the invoked configuration identify this new run independently of the 206-complex study. Loaded-weight hashes and the historical container digest were not recorded, so recovered file hashes are not runtime attestations.

The configuration requests 20 schedule points but executes 19 reverse updates, indexed 0--18. Four arms are copied from a shared parent immediately after native coordinate update 17. Each guided arm then receives one correction, the expanded batch performs native update 18, and each receives its second correction before confidence ranking. Composition-matched donors are chosen once from the post-update-17, pre-correction coordinates using the recorded element quotas and atom-index tie-break. The set is reused at update 18 and throughout both solver calls; a quota shortage raises an error rather than substituting another element.
\begin{table}[!htbp]
\centering
\caption{CICG intervention schedules for two frozen docking backbones. All update indices are zero based. Each correction call has at most five relinearizations.}
\label{tab:verified_hooks}
\fitbox{\begin{tabularx}{\linewidth}{@{}Yrrr@{}}\toprule
Configuration & Native updates & First correction & Second correction \\\midrule
SurfDock \citep{cao2025surfdock} + CICG (reference schedule)
& 0--19 & after 18 & after 19 \\
SurfDock \citep{cao2025surfdock} + terminal-only CICG
& 0--19 & after 19 & after first call \\
DiffDock-L \citep{corso2024diffdockl} + CICG
& 0--18 & after 17 & after 18 \\
\bottomrule\end{tabularx}}
\end{table}

All six pairwise arm combinations have matching retained pre-intervention hashes over 6,880 parents per pair. The configured final update has zero added noise, as confirmed by the recovered sampling code; no RNG-state or noise-tensor hashes were retained. Saved trace records include 55,040 arm--parent--hook rows. Of these, 13,760 are baseline no-correction events, 41,087 use the SVD operator, and 193 record unchanged-parent fallback for nonfinite or diverged parents. These are hook events, not independent pose failures.

All 172 planned query--seed blocks remain in the evaluation. In 153 blocks every one of the 160 arm/pose slots is parseable; the other blocks retain their invalid slots as failures. Invalid selected outputs are not replaced by lower-ranked valid candidates. Raw-file to parent linkage uses retained rankings and coordinate matching, including a record of nonunique assignments for identical coordinates. The retained final four-arm run, rather than incomplete earlier paired runs, supplies the comparisons.
\begin{table}[!htbp]
\centering
\caption{DiffDock-L follow-up on 86 queries, two seeds, and 40 planned poses per arm. Selected-pose counts use all 172 planned query--seed blocks. Invalid slots retain failure status.}
\label{tab:diffdock_followup_components}
\fitbox{\begin{tabularx}{\linewidth}{@{}Yrrrr@{}}\toprule
Condition & Joint & Ligand & Identity & Clash \\\midrule
No guidance & 43 & 84 & 82 & 5 \\
Historical donors & 60 & 80 & 97 & 8 \\
Composition-matched Nearest-$K$ & 59 & 81 & 97 & 9 \\
Native-donor oracle & 60 & 76 & 107 & 8 \\
\bottomrule\end{tabularx}}
\end{table}

Historical-minus-baseline is +9.88 points [3.09, 17.16], composition-matched-minus-baseline +9.30 [2.35, 17.16], and historical-minus-composition-matched +0.58 [$-2.03$, 3.52]. Historical and native guidance have identical joint labels on all 172 blocks; their ligand and realized-identity outcomes still differ. This agreement of observed binary outcomes is not equivalence of the policies or generated coordinates. Both backbones use the same biological queries, and no cross-backbone pooling increases the biological sample size.

\subsection{MetalloDock raw-pose operating evaluation}
\label{app:followup_metallodock}
The public MetalloDock checkpoint \citep{zhang2026metallodock} is run through a locally patched virtual-screening graph path with receptor-only metal-pocket inputs and independently embedded ETKDG ligand conformers. The pocket is a holo-derived 10~\AA\ crop around the crystal ligand. Thus native ligand coordinates are not supplied as ligand-pose inputs, but the pocket location is not a blind whole-protein prediction. Native donor labels and experimental ligand--metal bonds are absent from the model input; all 86 input graphs have zero ligand--metal graph edges. Existing bookkeeping and empty-edge patches mean this is not an unmodified upstream invocation.

After an eight-query technical pilot spanning six metal classes, the full run produces one raw pose for each of the 86 queries. The wrapper exports raw model coordinates and does not invoke OpenFF/OpenMM refinement. The checkpoint file is \texttt{docking\_\allowbreak metallodock\_\allowbreak time.pkl}; the reported and recovered-current hash agree. The historical invocation seed and loaded-weight attestation were not recorded, and a wrapper default is not treated as an executed argument. Per-query wrapper wall time totals 77.94~s with a median of 0.913~s; this excludes model loading and is not an equal-compute comparison.
\begin{table}[!htbp]
\centering
\caption{MetalloDock under the specified virtual-screening graph path. One raw pose is exported per query with refinement disabled. Pocket location is holo-derived; this is not a matched 40-pose or blind-pocket benchmark.}
\label{tab:metallodock_raw_followup}
\fitbox{\begin{tabularx}{\linewidth}{@{}Yrr@{}}\toprule
Outcome & Count & Rate (\%) \\\midrule
Raw pose generated & 86/86 & 100.00 \\
Ordinary Full & 39/86 & 45.35 \\
Clash-free joint success & 35/86 & 40.70 \\
Carbon stereochemistry inversion flagged & 20/86 & 23.26 \\
\bottomrule\end{tabularx}}
\end{table}

Indexed input/output/native graph parity was checked for all 86 queries. Raw coordinate reevaluation reproduces 39 Ordinary Full and 35 joint successes, with 20 carbon-stereochemistry inversion flags and 66 preserved inputs. The release includes atom mapping and center-level signed-volume records. No pose alignment or stereochemical repair is performed. These results describe this one-output, refinement-disabled path; the different information, refinement, and candidate budgets are retained in interpreting comparisons with the 40-pose banks.

\subsection{Paired resampling and selector provenance}
\label{app:followup_statistics}
The pooled comparisons use the recorded query-weighted, two-seed, 33-group analysis in Appendix~\ref{app:statistics}. The same 20,000 draws are applied to all 102 retained effect vectors; the ledger provides the paired sample IDs, per-query differences, group sums and denominators, and unrounded intervals. Replaying 48 earlier contrast-specific schedules recovers their historical intervals. For observations that are unchanged, small differences from earlier intervals arise from the recorded resampling streams and ordering; historical results remain in the provenance record.

\paragraph{Definition of the fixed-selector comparison.}
For one query--seed pair, write $\mathcal B_H=\mathcal B_{\widehat D}$ for the historical-guided bank. The same donor-aware selector independently returns
\[
\widehat y_0=S(\mathcal B_0;\widehat D),
\qquad
\widehat y_H=S(\mathcal B_H;\widehat D).
\]
The reported bank effect compares $R(\widehat y_H)$ with $R(\widehat y_0)$.

Let $k_0$ denote the recorded sample identifier of
$\widehat y_0$. A separate diagnostic instead evaluates
$R(y^H_{k_0})$, the pose in the guided bank with the
baseline-selected sample identifier, without independently reselecting a guided pose.

In the recovered records, reusing $k_0$ changes 87 historical selected sample identifiers and 14 joint outcomes (seven in each direction). Although both constructions total 51 baseline and 64 guided successes, their paired improvement/regression counts differ: 21/8 for independent bank selection and 17/4 for inherited sample identifiers. The reported fixed-selector result uses independent selection and its reproduced interval [1.11, 14.04]. The source snapshots and paired record comparisons are included with the analysis code.

\FloatBarrier
\label{page:appendixend}
\end{document}